\def\K{{\rm K}}

\def\muK{{\rm \mu K}}
\def\MJy{{\rm MJy}}
\def\Jy{{\rm Jy}}
\def\mJy{{\rm mJy}}
\def\sr{{\rm sr}}
\def\MJysr{\MJy/\sr}

\def\GHz{{\rm GHz}}

\def\ra{{\rm ra}}
\def\dec{{\rm dec}}

\def\expec#1{\langle#1\rangle}
\def\bigexpec#1{\left\langle#1\right\rangle}
\def\tento#1{\times 10^{#1}}

\def\etal{{\frenchspacing\it et al.}}
\def\ie{{\frenchspacing\it i.e.}}
\def\eg{{\frenchspacing\it e.g.}}
\def\etc{{\frenchspacing\it etc.}}
\def\rms{{\frenchspacing r.m.s.}}

\def\pp{\noindent\parshape 2 0truecm 13.6truecm 1truecm 12.6truecm}
\def\rf#1;#2;#3;#4 {\par\pp#1, {\it #2}, {\bf #3}, #4. \par}
\def\rn{\pp}

\def\beq#1{\begin{equation}\label{#1}}
\def\eeq{\end{equation}}
\def\beqa#1{\begin{eqnarray}\label{#1}}
\def\eeqa{\end{eqnarray}}
\def\eq#1{equation~(\ref{#1})}
\def\Eq#1{Equation~(\ref{#1})}
\def\eqnum#1{~(\ref{#1})}

\def\bfig{\begin{figure}[h] \centerline{\hbox{}}\vfill}
\def\efig{\end{figure}\vfill\newpage}

\def\fig#1{Figure~\ref{#1}}

\def\spose#1{\hbox to 0pt{#1\hss}}
\def\simlt{\mathrel{\spose{\lower 3pt\hbox{$\mathchar"218$}}
     \raise 2.0pt\hbox{$\mathchar"13C$}}}
\def\simgt{\mathrel{\spose{\lower 3pt\hbox{$\mathchar"218$}}
     \raise 2.0pt\hbox{$\mathchar"13E$}}}
\def\simpropto{\mathrel{\spose{\lower 3pt\hbox{$\mathchar"218$}}
     \raise 2.0pt\hbox{$\propto$}}}

\def\addr#1{{\small\it #1}}
\def\auth#1{{#1}}

\def\vx{{\bf x}}

\def\va{{\bf a}}

\def\va{{\bf a}}
\def\N{N}

\def\l{\ell}

\def\expec#1{\langle#1\rangle}

\def\Pn{P^{(n)}}
\def\Ps{P^{(s)}}

\def\vr{{\bf r}}
\def\rh{\widehat{\bf r}}

\def\FWHM{{\rm FWHM}}
\def\ignore#1{}

\def\ed{\end{document}}

\documentstyle[11pt,epsf]{article}
\begin{document}
%\baselineskip=0.3333truein
%\baselineskip=20pt

%%%%%%%%%%%%%%%%%%%%%%%%%%%%%

\begin{titlepage}   % Not numbered.

\noindent
%\today
\hfill MPI-PhT/95-62
\begin{center}

\vskip0.9truecm
{\bf
A METHOD FOR SUBTRACTING 
FOREGROUNDS FROM MULTI-FREQUENCY CMB SKY MAPS
\footnote{Published in {\it MNRAS}, {\bf 281}, 1297 (1996).
First submitted June 19, 1995.\\
Available from 
{\it h t t p://www.sns.ias.edu/$\tilde{~}$max/wiener.html} (faster from the US)\\
and from
{\it h t t p://www.mpa-garching.mpg.de/$\tilde{~}$max/wiener.html} (faster from Europe).\\
Color figures are also available on these web pages.}
}

\vskip 0.5truecm
  \auth{Max Tegmark}
  \smallskip

  \addr{Max-Planck-Institut f\"ur Physik, F\"ohringer Ring 6,}
  \addr{D-80805 M\"unchen;}

 \addr{email: max@mppmu.mpg.de}
 
  \bigskip
  \auth{George Efstathiou}\\
  \smallskip
  \addr{Dept. of Physics, University of Oxford, Keeble Road, Oxford, OX1 3RH;}

 \addr{email: g.efstathiou@physics.oxford.ac.uk}
  \smallskip
  \vskip 0.8truecm

{\bf Abstract}
\end{center}
\bigskip
An improved method for subtracting contaminants from Cosmic Microwave
Background (CMB) sky maps is presented, and used to estimate how well
future experiments will be able to recover the primordial CMB
fluctuations.  We find that the naive method of subtracting
foregrounds (such as dust emission, synchrotron radiation,
free-free-emission, unresolved point sources, {\etc}) on a pixel by
pixel basis can be improved by more than an order of magnitude by
taking advantage of the correlation 
of the emission in neighboring pixels. 
The optimal multi-frequency 
subtraction method improves on simple pixel-by-pixel
subtraction both by taking noise-levels into account, and
by exploiting the fact that most contaminants have angular
power spectra that differ substantially from that of the CMB. The
results are natural to visualize in the two-dimensional plane with
axes defined by multipole $\ell$ and frequency $\nu$. We present a
brief overview of the geography of this plane, showing the regions
probed by various experiments and where we expect contaminants
to  dominate. We
illustrate the method by estimating how well the proposed ESA
COBRAS/SAMBA mission will be able to recover the CMB fluctuations
against contaminating foregrounds.

\end{titlepage}

%%%%%%%%%%%%%%%%%%%%%%%%%%%%%%%%%%%%%%%%%%%%%%

\def\F{F}
\def\G{G}
\def\N{N}
\def\S{S}
\def\W{W}
\def\vk{{\bf k}}
\def\vr{{\bf r}}
\def\vx{{\bf x}}
\def\vy{{\bf y}}
\def\xr{x'}
\def\vxr{{\bf x}'}
\def\diag{{\rm diag}}

% Boldface Greek letters:
\font\bfmath=cmmib10
\def\bfvarepsilon{\hbox{\bfmath\char'042}}

\def\noise{\varepsilon}
\def\vnoise{\bfvarepsilon}
\def\noiseh{\widehat{\noise}}
\def\vnoiseh{\widehat{\vnoise}}
\def\error{\Delta}
\def\verror{{\bf\error}}
\def\errorh{\widehat{\error}}
\def\verrorh{\widehat{\verror}}
\def\xh{\widehat{x}}
\def\yh{\widehat{y}}
\def\vxh{\widehat{\vx}}
\def\vyh{\widehat{\vy}}
\def\vxrh{\widehat{\vxr}}
\def\xrh{\widehat{\xr}}
\def\Wh{\widehat{\W}}
\def\notyet{\vskip1cm{\Large [NOT WRITTEN YET]}\vskip1cm}

\section{INTRODUCTION}

There has been a surge of interest in the cosmic microwave
background radiation (CMB) 
since the first
anisotropies of assumed cosmological origin were
detected by the COBE DMR experiment (Smoot {\etal} 1992).
On the experimental front, many new experiments have been
carried out, and more are planned or proposed for
the near future (see White {\etal} 1994, for a review).
On the theoretical front, considerable progress
has been made in understanding
how the CMB power spectrum $C_\l$ depends on various
cosmological model parameters (see 
Bond {\etal} 1994, Hu \& Sugiyama 1995,
and references therein for recent reviews of
analytical and quantitative aspects of this problem).
It is now fairly clear that an accurate measurement 
of the angular power spectrum  $C_\l$ to multipoles
$\l \sim 10^3$ could  provide accurate constraints on many of 
the standard cosmological parameters ($\Omega$, $\Omega_b$, 
$\Lambda$, spectral index $n$, {\etc}), 
thus becoming the definitive arbiter between various
flavours of the cold dark matter (CDM) cosmogony and
other theories of the origin of structure
in the Universe.
To accurately measure the power spectrum and reach this goal, 
a number of hurdles must be overcome:
\begin{itemize}
\item Technical problems
\item Incomplete sky-coverage
\item Foregrounds
\end{itemize}
There is of course a wide variety of technical challenges that must be
tackled to attain high resolution, low noise, well-calibrated
temperature data over a wide range of frequencies and over most of the
sky. However, thanks to the rapid advance in detector technology over
the last two decades and the possibilities of ground based
interferometers, long duration balloon flights and space-born
experiments, there is a real prospect that high sensitivity maps of
the CMB will be obtained within the next decade.

The second hurdle refers to the fact that we are unlikely to 
measure the primordial CMB sky accurately behind the Galactic plane.
As is well known, the resulting incomplete sky coverage makes it
impossible to exactly recover the spherical harmonic coefficients 
that would give direct estimates of the angular power spectrum
$C_\l$.
However, a number of methods for efficiently constraining models, using
only partial sky coverage,  have now been developed
(G\'orski 1994; Bond 1995; Bunn \& Sugiyama 1995; Tegmark \& Bunn 1995), 
and it has recently been shown (Tegmark 1995, hereafter T96)
that even the individual $C_\l$-coefficients
can be accurately estimated for all but the
very lowest multipoles such as $\l=2$, by 
expanding the data in
appropriately chosen basis functions. 
Rather, the most difficult of the above-mentioned hurdles 
appears to be the third,
which is the topic of the present paper.

\begin{figure}[phbt]
\centerline{\epsfxsize=17cm\epsfbox{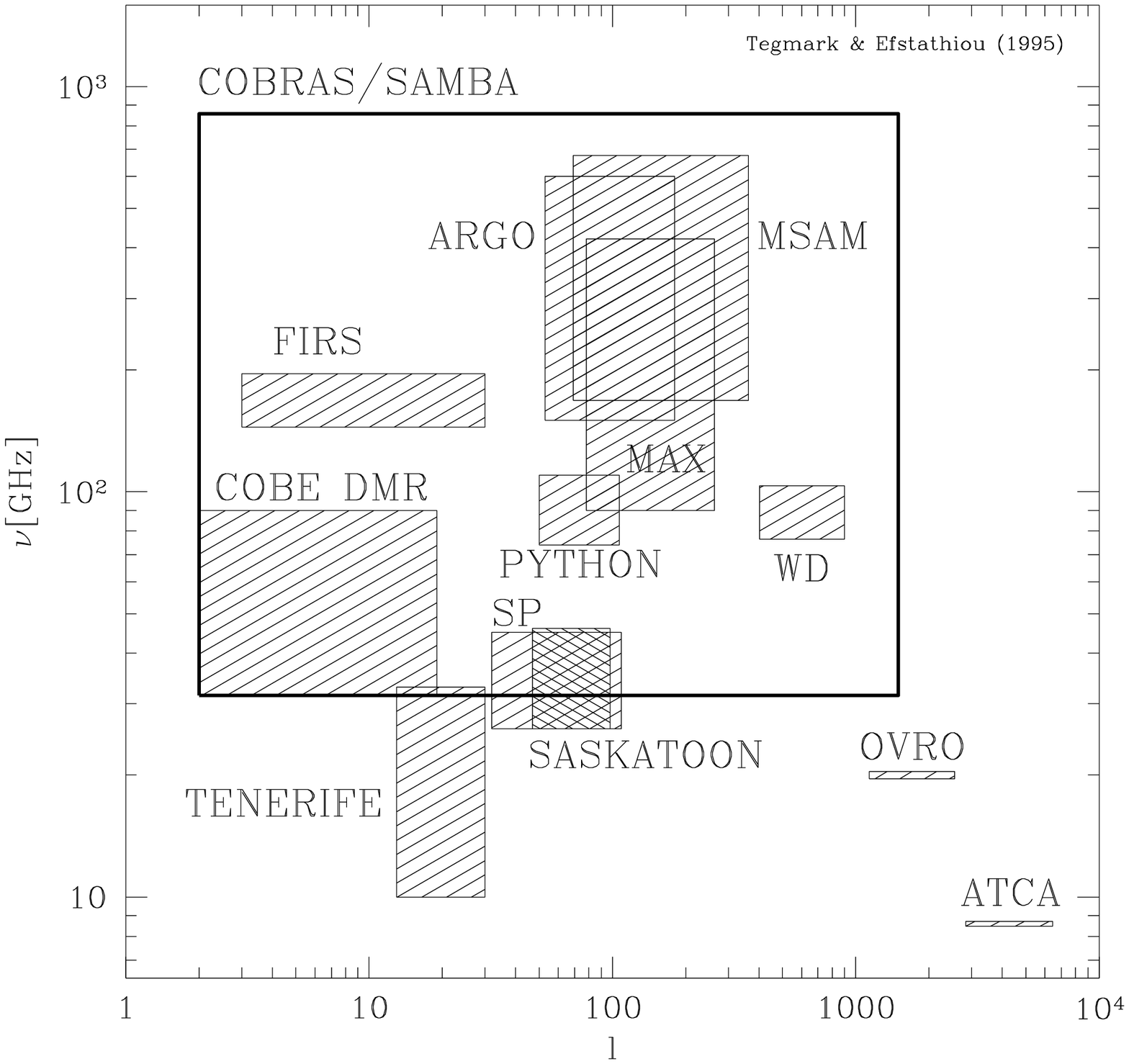}}
\caption{Where various CMB experiments are sensitive.}
The boxes roughly indicate the range of multipoles $\l$ and frequencies
$\nu$ probed by various CMB experiments. The large heavy unshaded box 
corresponds to the proposed COBRAS/SAMBA satellite.
\label{ExperimentsFig}
\end{figure}

The frequency-dependence of the various 
foregrounds has been extensively studied, in both 
``clean" and ``dirty" regions
of the sky (see {\eg} Brandt {\etal} 1993,
Toffolatti {\etal} 1994, for recent reviews).
However, these properties alone 
provide a description of the foregrounds
that is somewhat too crude to assess the extent
to which they can be separated from the underlying CMB signal,
since the foregrounds fluctuations depend on the multipole 
moment $\l$ as well (see 
Bouchet {\etal}, 1995, for  simulations). 
Most published plots comparing different CMB
experiments tend to show  $\l$ on the 
horizontal axis and an amplitude ($C_\ell$ or an {\rms}
$\Delta T/T$) on the vertical axis, whereas  most plots comparing 
different foregrounds show amplitude plotted against
frequency $\nu$.
Since the fluctuations in the latter tend to depend
strongly on both $\l$ and $\nu$, {\ie}, on both spatial 
and temporal frequency, 
one obtains a more accurate picture by combining both of these 
pieces of information and working in a 
two-dimensional plane as in Figures 1-6. We will indeed
find that the $\l$--$\nu$
plane arises naturally in the optimized subtraction
scheme that we present.
\fig{ExperimentsFig} shows roughly the regions in this plane 
probed by various CMB experiments.
Each rectangle corresponds to one experiment. Its 
extent in the $\l$-direction shows the customary
$1\sigma$ width of the experimental window function
(see {\it e.g.} White \& Srednicki 1995),
whereas the vertical extent shows the frequency range
that is covered. For single-channel experiments, we
plot the quoted bandwidth, whereas for multi-channel 
experiments, the box has simply been plotted in 
the range between the lowest and highest frequency channel.
For a more detailed description of these experiments, see
Scott, Silk \& White (1995) and references therein. 
Figures 2 though 5, which will be described in 
Section~\ref{ForegroundsSec},
show the estimated fluctuations of the CMB
and various foregrounds in the same plane, and comparing these
figures with \fig{ExperimentsFig} as in \fig{EverythingFig}, 
it is easy to understand which experiments are the most 
affected by the various foregrounds.
Moreover, as we will see, familiarity with the geography 
of this plane provides an intuitive understanding of 
the advantages and shortcomings of
different methods of foreground subtraction. 

A number of CMB satellite missions are currently under consideration
by various funding agencies, which would offer the excellent
sensitivity, resolution and frequency coverage that is needed for
accurate foreground subtraction.  Throughout this paper, we will use
the proposed European Space Agency COBRAS/SAMBA mission 
(Mandolesi {\etal} 1995) as an illustration of what the next 
generation of space-borne CMB experiments may be able to
achieve.  The specifications of 
competing satellite proposals are similar, though with 
a more restricted frequency range.

The rest of this paper is organized as follows.  After establishing
our basic notation in Section~\ref{NotationSec}, we derive the optimized
multi-frequency subtraction method in Section~\ref{WienerSec}.
In Section~\ref{ForegroundsSec}, we estimate the angular power spectra
of the various foreground contaminants.  In Section~\ref{ResultsSec},
we use these estimates to assess the effectiveness of the subtraction
technique and to show how accurately the CMB fluctuations could be
recovered from high quality data, such as might be obtained from the
proposed COBRAS/SAMBA satellite.

\section{NOTATION}
\label{NotationSec}

\def\Bdust{B_{dust}}
\def\Bsynch{B_{synch}}
\def\Bff{B_{ff}}
\def\Bps{B_{ps}}
\def\Bgal{B_{gal}}
\def\Bsz{B_{sz}}
\def\Bcmb{B_{cmb}}
\def\Bplanck{B_0}
\def\Bnoise{B_{noise}}
\def\dT{\delta T}
\def\dTdust{\delta T_{dust}}
\def\dTsynch{\delta T_{synch}}
\def\dTff{\delta T_{ff}}
\def\dTps{\delta T_{ps}}
\def\dTgal{\delta T_{gal}}
\def\dTsz{\delta T_{sz}}
\def\dTcmb{\delta T_{cmb}}
\def\dTnoise{\delta T_{noise}}
\def\Cnoise{C^{noise}}
\def\Cdust{C^{dust}}
\def\Cps{C^{ps}}
\def\anoise{a^{noise}}

\def\alm{a_{\l m}}
\def\almp{a_{\l' m'}}

\begin{figure}[phbt]
\centerline{\epsfxsize=17cm\epsfbox{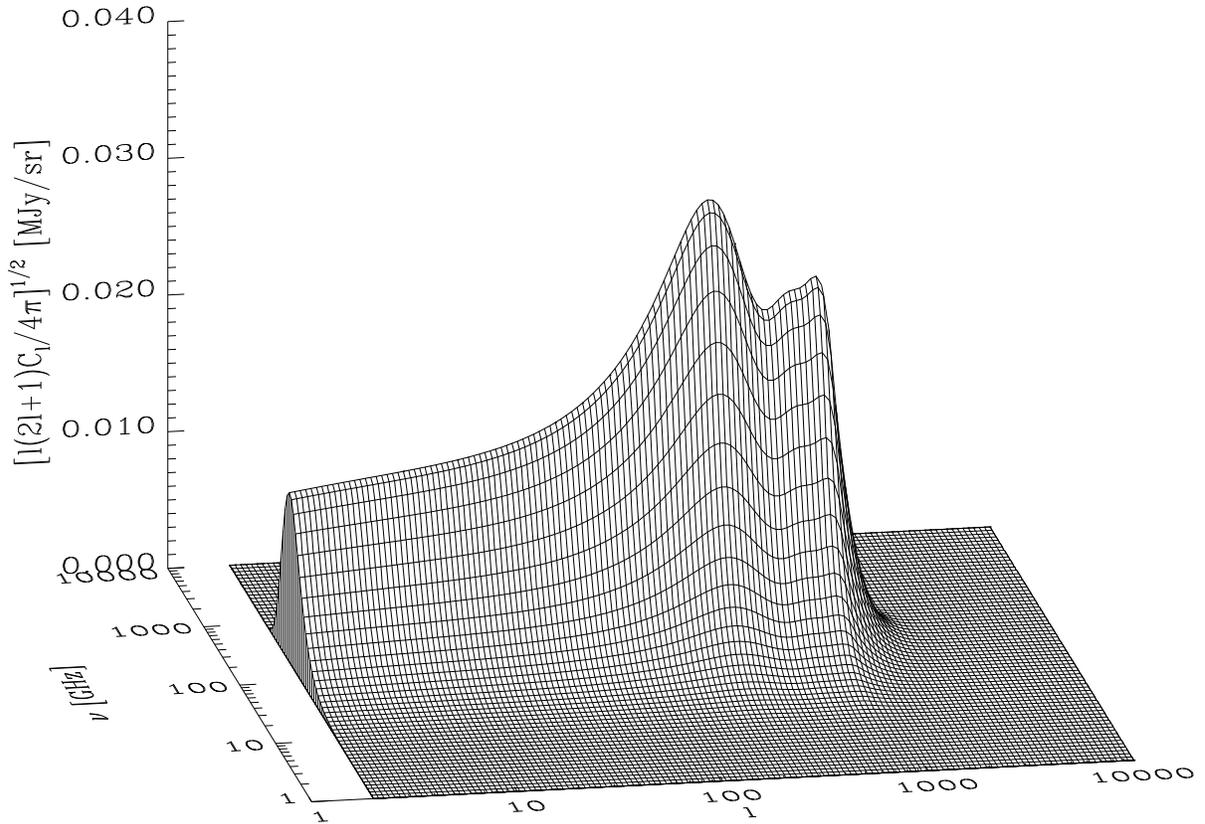}}
\caption{How the CMB brightness fluctuations depend 
on multipole and frequency in the standard CDM model
(assuming scale-invariant scalar 
fluctuations in a critical density, $\Omega=1$,
universe, with Hubble constant $H_0 = 50{\rm km}{\rm s}^{-1}{\rm
Mpc}^{-1}$, and baryon density $\Omega_b = 0.05$). The CDM power
spectrum was computed as described by Bond \& Efstathiou (1987).}
\label{CMB_B_Fig}
\end{figure}

\begin{figure}[phbt]
\centerline{\epsfxsize=17cm\epsfbox{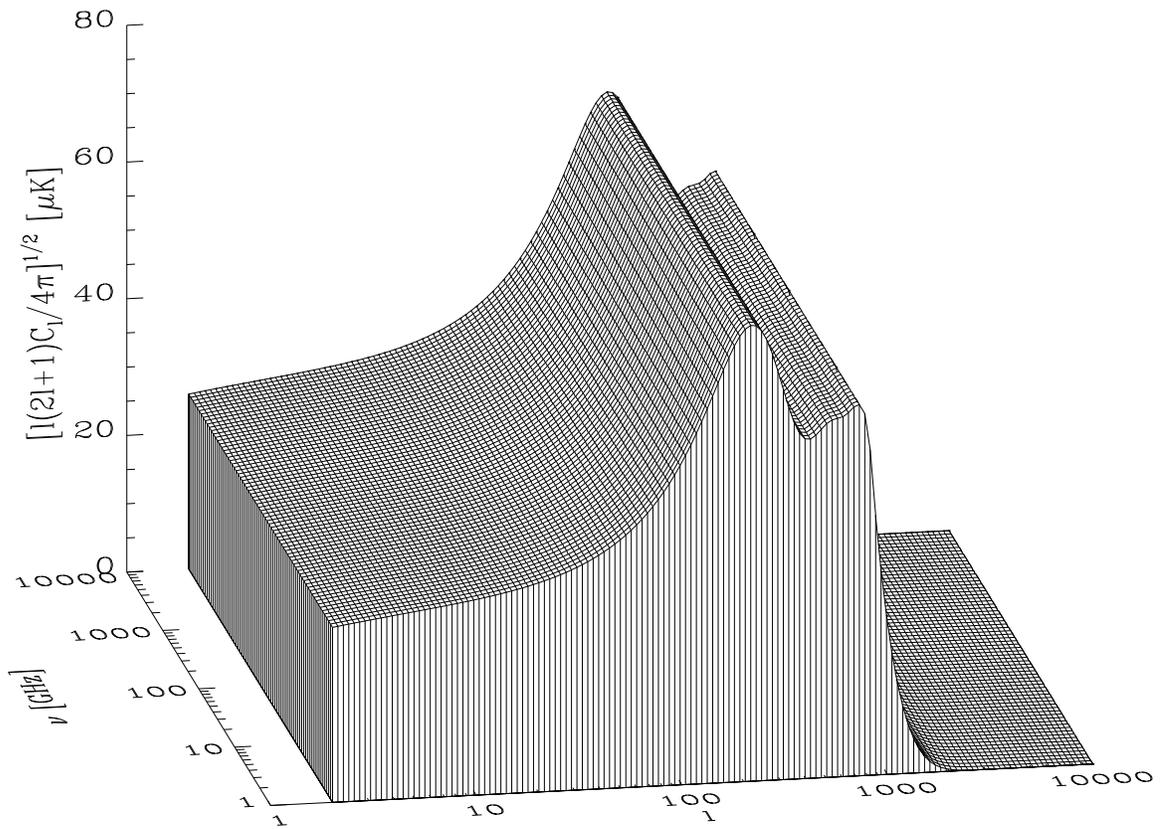}}
\caption{How the CMB temperature fluctuations depend 
on multipole and frequency for the CDM model with parameters
as in Figure 2.}
\label{CMB_T_Fig}
\end{figure}

Let $B(\rh,\nu)$ denote the total sky brightness at frequency 
$\nu$ in the direction of the unit vector $\rh$.
Since we know that $B$ is a sum of contributions of physically
distinct origins, we will write it as
\beq{BsumEq}
B(\rh,\nu) = \sum B_i(\rh,\nu).
\eeq
In the microwave part of the spectrum, the most important 
non-CMB components
are synchrotron radiation $\Bsynch$, free-free emission
$\Bff$, dust emission $\Bdust$, and radiation from point sources
$\Bps$ (both radio sources and infrared emission from
galaxies). We will separate the CMB-contribution into two
terms; an isotropic blackbody $\Bplanck$ and the fluctuations 
$\Bcmb$ around it. The former, which is of course independent of
$\rh$, is given by
\beq{B0eq}
\Bplanck(\nu) = {2h\over c^2}{\nu^3\over e^x-1} = 
2{(kT_0)^3\over (hc)^2}\left({x^3\over e^x-1}\right)
\approx 270.2\,\MJysr \left({x^3\over e^x-1}\right),
\eeq
where $x\equiv h\nu/kT_0\approx \nu/56.8\GHz$ and $T_0\approx 2.726\K$ 
(Mather {\etal} 1994).
If the actual CMB temperature across the sky is 
$T(\rh) = T_0 +\delta T(\rh)$, then 
to an excellent approximation, 
$\Bcmb(\rh,\nu) = (\partial\Bplanck(\nu)/\partial T_0)\delta T(\rh)$.
(The quadratic correction will be down by a factor of
$\delta T/T\approx 10^{-5}$.)
This conversion factor from brightness to temperature is 
\beq{dBdTeq}
{\partial\Bplanck\over\partial T_0} = 
{k\over 2} \left({k T_0\over hc}\right)^2
\left({x^2\over\sinh x/2}\right)^2
\approx
\left({24.8\MJy/\sr\over\K}\right)
\left({x^2\over\sinh x/2}\right)^2.
\eeq
Since any other contaminants that we fail to take into account
will be mistaken for CMB fluctuations, it is convenient to 
convert all brightness fluctuations into temperature fluctuations
in the analogous way, so we define
\beq{ConversionEq1}
\dT_i(\rh,\nu) \equiv
{B_i(\rh,\nu)\over 
\partial\Bplanck/\partial T_0}.
\eeq
Note that with this definition, $\dTcmb(\rh,\nu)$ is independent
of $\nu$ (compare \fig{CMB_B_Fig} with \fig{CMB_T_Fig}), 
whereas most of the foregrounds will exhibit a 
strong frequency dependence.
We expand the temperature fluctuations in spherical harmonics as
usual;
\beq{MultipoleExpansionEq}
\dT_i(\rh,\nu) = 
\sum_{\l=0}^{\infty}
\sum_{m=-\l}^{\l}
Y_{\l m}(\rh) \,\alm^{(i)}(\nu).
\eeq
For isotropic CMB fluctuations, we have
\beqa{CMBexpecEq1}
\expec{\alm^{cmb}(\nu)}&=&0,\\
\label{CMBexpecEq2}
\expec{\alm^{cmb}(\nu)^* \almp^{cmb}(\nu)}&=&
\delta_{\l\l'}\delta_{m m'} C_\l,
\eeqa
where $C_\l$ is the  frequency independent 
{\it angular power spectrum}.
For other components, the means
$\expec{\alm^{(i)}(\nu)}$ are not necessarily equal to
zero. For instance, most of the foregrounds are
by nature non-negative ($B_i \ge 0)$, so  we expect the
monopole to be positive. 
Also, if there are deviations from isotropy (a secant behavior
with Galactic latitude, for instance), we will not have
$\expec{\alm^{(i)}(\nu)^* \almp^{(i)}(\nu)} \propto
\delta_{\l\l'}\delta_{m m'}$.
For such cases, we simply define 
\beq{UglyCldefEq}
C_\l^{(i)}(\nu) \equiv 
{1\over 2\l+1}\sum_{m=-\l}^\l \bigexpec{\left|\alm^{(i)}(\nu)\right|^2},
\eeq
since this is the quantity that on average will be added to the estimate of the
CMB power spectrum if the contaminant is not removed.
As we shall see further on, all contaminants that we 
have investigated in this paper do  become
fairly isotropic when we mask out all but the cleanest parts of the
sky; the different Fourier components do indeed decouple and
so the only important difference from CMB behavior is an additional 
constant term in the monopole (which is, of course,
unmeasurable anyway).

We conclude this Section with a few comments on how to 
read Figures 2 through 6.
If a random field satisfies equations\eqnum{CMBexpecEq1}
and\eqnum{CMBexpecEq2}, the addition theorem for 
spherical harmonics gives the well-known result
\beq{rmsEq}
\expec{\dT_i(\nu)^2} = \sum_{l=0}^{\infty}
\left({2\l + 1\over 4\pi}\right) C_\l^{(i)}(\nu) 
\approx \int \left[{\l(2\l + 1)\over 4\pi}\right] C_\l^{(i)} d(\ln\l).
\eeq
This is the reason that we plot the quantity
$[\l(2\l + 1)C_l^{(i)}(\nu)/4\pi]^{1/2}$ in the
figures of the next section: the resulting {\rms} temperature
fluctuation $\dT_i(\nu)$ is basically just the {\rms} 
height of the curve times a small constant.
If we compute the {\rms} average in a multipole range 
$\l_0\leq\l\leq\l_1$, then 
this constant is simply $[(\ln(\l_1/\l_0)]^{1/2}$.
For the CMB temperature 
fluctuations of standard CDM, for instance, 
which are shown in \fig{CMB_T_Fig} normalized to COBE
(Bunn {\etal} 1995),
$\expec{\dT^2}^{1/2} \approx 108\muK$.
Convolving the fluctuations with the COBE beam with $\FWHM=7.08^{\circ}$
suppresses $C_\l$ for $\l\gg 10$, giving
$\expec{\dT^2}^{1/2} \approx 37\muK$. The order of magnitude of
both of these numbers can be roughly read off by eye with the above
prescription.
For the CMB brightness fluctuations 
shown in \fig{CMB_B_Fig}, everything is completely analogous.
For instance, the total fluctuations are
$\expec{\Bcmb(\nu)^2}^{1/2} \approx 0.05\MJysr$ at the
maximum sensitivity frequency $\nu\approx 218\>\GHz$,
and $0.02\MJysr$ as seen with the COBE beam.

\section{MULTI-FREQUENCY FOREGROUND SUBTRACTION}
\label{WienerSec}

In this Section, we
derive the multi-frequency subtraction scheme mentioned in
the introduction. Given data in several frequency channels, the goal
is to produce the best maps corresponding to 
the different physical components, where ``best" is 
taken to mean having
the smallest {\rms} errors.
We first derive the method for an idealized case, and then 
add the various real-world complications one by one. 

Before beginning, it is instructive to compare this 
approach with that of likelihood analysis.
Most published analyses of the COBE DMR sky maps 
(see {\it e.g.} Tegmark \& Bunn, 1995 for a recent review) have used
likelihood techniques to constrain models with power spectra 
described by one or two 
free parameters, typically a spectral index and an 
overall normalization.
As long as the number of model-parameters is rather small, 
a useful way to deal with foreground contamination is that
described by
Dodelson \& Stebbins (1994). The basic idea is 
to include in the likelihood analysis a number of
``nuisance parameters" describing the foregrounds,
and then marginalize over these parameters to obtain the 
Bayesian probability distribution for the parameters of interest.
Despite its elegance, this method is of course only feasible when the number
of parameters, $n$, to be estimated is small, since the number 
of grid points in the $n$-dimensional parameter space
(and hence the amount of computer time required for the analysis)
grows exponentially with $n$. 
The problems addressed in this paper are how to 
estimate the entire power spectrum $C_\l$
(about $n=10^3$ parameters) and how to 
reconstruct a high-resolution all-sky map (with perhaps as 
many as $n\sim 10^7$ parameters (pixels)),
which is why a more direct approach other than likelihood analysis
is required.

The method we present below is type of linear filtering closely
related to Wiener filtering, but with a crucial difference.
Linear filtering techniques have recently been applied to a 
range of cosmological problems. Rybicki \& Press (1992) give a detailed 
discussion of the one-dimensional problem. 
Lahav {\etal} (1994), Fisher {\etal} (1995) and Zaroubi {\etal} (1995)
apply Wiener filtering to galaxy surveys. 
In particular, Bunn {\etal} (1994) apply Wiener filtering to the 
COBE DMR maps. We will generalize this treatment to the case of multiple 
frequencies and more than one physical component.
Although it is tempting to 
refer to this simply as multi-frequency Wiener filtering, 
we will avoid this term, as it can cause confusion for the following
reason.
As is discussed in any signal-processing text, regular 
Wiener filtering modifies the power spectrum of the signal. 
Specifically, the filtered signal will have a power spectrum 
\beq{StandardWienerEq}
P'_s(k) = {P_s(k)\over P_s(k)+P_n(k)},
\eeq
where $P_s$ and $P_n$ are the power spectra of the true signal and the 
contaminant, respectively.
This of course makes it useless for power-spectrum estimation. 
As we shall see in 
Section \ref{RenormSec}, our approach {\it does not} alter the
power spectrum of the signal (the CMB, say), and
moreover has the attractive property of 
being independent of any assumptions about
the true CMB power spectrum, requiring only assumptions about
the power spectra of the foregrounds.
This is possible because more than one frequency channel is available,
which allows foreground/background separation even if the
two have identical power spectra. 
The availability of multiple frequencies is 
absolutely essential 
to our method:
in the special 
case where $m$, the number of channels, equals one, 
it degenerates not to standard Wiener filtering,
but to the trivial case of doing no subtraction at all.

The method that we advocate turns out to be related to 
Wiener filtering by a simple $\l$-dependent rescaling of 
the weights. It is therefore trivial switch
between this power-conserving subtraction scheme 
and an error-minimizing multi-frequency 
generalization of Wiener filtering. 
For pedagogical purposes, we first present the latter,
in sections~\ref{FlatSec}-\ref{NonGaussSec},
then show how it should 
be rescaled in Section~\ref{RenormSec}.

\subsection{The model}
\label{ModelSec}

Let us make the approximation that we can can write
\beq{ComponentSumEq}
{\delta T}(\vr,\nu) = 
\sum_{i=1}^n f_i(\nu) x_i(\vr),
\eeq
where each term corresponds to a distinct physical component
(such as CMB, dust, synchrotron radiation, free-free emission, 
radio point sources, {\etc}).
Thus we are simply assuming that the contribution 
from each component is separable into a function 
that depends only on frequency times a function 
that depends only on position.
For definiteness, let us normalize all
the functions $f_i$ so that $f_i(100\,\GHz)=1$, thus absorbing 
the physical units into the fields $x_i$. 

Suppose that we observe the sky in $m$ different frequency
channels, so that at each point $\vr$ on the sky, we measure 
$m$ different numbers $y_i(\vr)$, $i=1,...,m$.
(These need of course not be different channels observed by 
the same experiment  --- we may for instance want to 
use the IRAS 100 micron survey as an additional ``channel".)
Assuming that we know the spectra $f_i(\nu)$ of
all the components,
we can write
\beq{ModelEq}
y(\vr) = F x(\vr) + \vnoise(\vr).
\eeq
Here the vector $\vnoise(\vr)$ corresponds to the instrumental noise
in the various channels,
and $F$ is a fixed $m\times n$ matrix, the {\it frequency response matrix},
given by
\beq{FdefEq}
F_{ij} \equiv \int_0^{\infty} w_i(\nu) f_j(\nu) d\nu,
\eeq
$w_i(\nu)$ being the frequency response of the $i^{th}$ channel.

\subsection{The idealized flat case}
\label{FlatSec}

We now turn to the highly idealized case 
where the sky is flat rather than spherical, 
there is no pixelization, no Galactic zone of avoidance, {\etc}
This simple case is directly applicable only to a 
small  patch of sky sampled at high resolution.
In the subsequent sections, we will show how to tackle
the numerous real-world complications.
It will be seen that none of these complications
change the basic matrix prescription of the simple case
described here.

The second term in \eq{ModelEq}, 
the vector $\vnoise$, contains the instrumental
noise in the different  frequency channels. We model this by
\beqa{NoiseDefEq1}
\expec{\noise_i(\vr)}&=& 0,\\
\label{NoiseDefEq2}
\expec{\noiseh_i(\vk)^*\noiseh_j(\vk')} &=& 
(2\pi)^2 \delta_{ij} \delta(\vk'-\vk) \Pn_i(k),
\eeqa
thereby allowing for the possibility that the noise within
each channel may exhibit some 
correlation (hats denote Fourier transforms).
Uncorrelated noise simply corresponds to the case
where the noise power spectra are given by $\Pn_i(\vk) = \sigma_i^2$,
where the $\sigma_i$ are constants. 

Analogously, we assume that the physical components satisfy
\beqa{SignalDefEq1}
\expec{x_i(\vr)}&=& 0,\\
\label{SignalDefEq2}
\expec{\xh_i(\vk)^*\xh_j(\vk')} &=&
(2\pi)^2\delta_{ij} \delta(\vk'-\vk) \Ps_i(k).
\eeqa
Note that we are not assuming the random fields to be Gaussian.
\Eq{SignalDefEq2} follows directly from \eq{SignalDefEq1} and
homogeneity (translational invariance), together with the obvious assumption 
that the different components are independent.

Our goal is to make a reconstruction, denoted $\vx'$, 
of the physical fields $\vx$ from the 
observed data $\vy$. We will set
out to find the best {\it linear} reconstruction.
Because of the translational invariance,
the most general linear estimate $\vxr$ of $\vx$ can clearly 
be written as
\beq{WdefEq}
\vxr(\vr) = \int W(\vr-\vr')\vy(\vr')d^2 r',
\eeq
for some matrix-valued function $W$ that we will refer to as the
{\it reconstruction matrix}. 
We now proceed to derive the best choice of the reconstruction matrix.
Defining the {\it reconstruction errors} as
$\error_i(\vr)\equiv \xr_i(\vr)-x_i(\vr)$, a straightforward calculation
gives
\beqa{ErrorEq1}
\expec{\error_i(\vr)}&=& 0,\\
\label{ErrorEq2}
\expec{\error_i(\vr)^2} &=& 
\int\Bigg[\sum_{j=1}^n\left|\left(\Wh(\vk)\F-I\right)_{ij}\right|^2\Ps_j(\vk)
 \nonumber \\
 & & + \sum_{j=1}^m\left|\Wh_{ij}(\vk)\right|^2\Pn_j(\vk)
\Bigg]\;d^2k, \eeqa
independent of $\vr$.
\Eq{ErrorEq1} merely tells us that our estimators of the fields
are unbiased. Pursuing the analogy of ordinary Wiener filtering, we 
select the reconstruction matrix 
that minimizes the {\rms} errors, {\ie}, minimizes
$\expec{\error_i(\vr)^2}$.
We thus require $\delta \expec{\error_i(\vr)^2} = 0$, where the variation is 
carried out with respect to $\Wh_{ij}$, and obtain
\beq{MessySolutionEq}
\sum_{k=1}^n\left(\Wh(\vk)\F-I\right)_{ik}F_{jk}\Ps_k(\vk)
+
\sum_{k=1}^m \Wh_{ik}(\vk)\Pn_k(\vk)\delta_{kj} = 0.
\eeq
Defining the noise and signal matrices $N$ and $S$ by 
\beqa{GdefEq}
%G_{ij}(\vk) &\equiv&\sum_{k=1}^n \F_{ik}\F_{jk}\Ps_k(\vk),\\
N_{ij}(\vk) &\equiv&\delta_{ij}\Pn_i(\vk) = \diag\{\Pn_i(\vk)\},\\
S_{ij}(\vk) &\equiv&\delta_{ij}\Ps_i(\vk) = \diag\{\Ps_i(\vk)\},
\eeqa
\eq{MessySolutionEq} reduces to simply $\Wh[FSF^t+N]-SF^t=0$, 
which has the solution
\beq{WienerResultEq}
\Wh(\vk) = S(\vk) F^t [FS(\vk)F^t + N(\vk)]^{-1}.
\eeq
These are the appropriate formulae to use when limiting the attention to a 
rectangular patch of sky whose sides are  small enough
($\ll$ one radian $\approx 60^\circ$) that its curvature can be neglected.
With all sky coverage, we need to solve the corresponding subtraction 
problem on a sphere instead, which is the topic of the next subsection.

\subsection{The idealized spherical case}

Above we saw that the optized subtraction became much simpler in Fourier space,
where it became diagonal. In other words, although the linear combination 
mixed the $m$ different frequencies, it never mixed
Fourier coefficients corresponding to different wave vectors $\vk$. 
The generalization of the derivation above to the case of fields on the
celestial sphere is trivial, and not surprisingly, the corresponding natural
basis functions in which the subtraction becomes diagonal are the spherical
harmonics. Expanding all fields in spherical harmonics as in 
Section~\ref{NotationSec}, and combining the 
observed $a_{\l m}$-coefficients for the various frequency channels 
in the vector 
$\va_{\l m}$, we can thus write our estimate of the true
coefficients for the various components, denoted $\va'_{\l m}$, as
\beq{SphWdefEq}
\va'_{\l m} = W^{(\l)} \va_{\l m}.
\eeq
The analogues of equations\eqnum{GdefEq}-(\ref{WienerResultEq}), 
giving the reconstruction matrix $W^{(\l)}$, become
\beqa{SphWienerResultEq}
\W^{(\l)}&=&S^{(\l)} F^t [FS^{(\l)}F^t + N^{(\l)}]^{-1},\\
%G^{(\l)}_{ij}&\equiv&\sum_{k=1}^n \F_{ik}\F_{jk}C_\l^{(k)},\\
N^{(\l)}_{ij}&\equiv&\delta_{ij}C_\l^{noise,i},\\
S^{(\l)}_{ij}&\equiv&\delta_{ij}C_\l^{(i)}.
\eeqa
The corresponding reconstruction errors 
$\Delta {a'}^{(i)}_{lm}$
have 
their mean square value $\Delta C_\l^{(i)}\equiv
\expec{|\Delta {a'}^{(i)}_{lm}|^2}$ given by
\beq{SphErrorEq}
\Delta C_\l^{(i)} = 
\left[
\sum_{j=1}^n\left|\left(W^{(\l)}\F-I\right)_{ij}\right|^2 C_\l^{(j)}
+
\sum_{j=1}^m\left|\W^{(\l)}_{ij}\right|^2 C_\l^{noise,j}
\right].
\eeq
As we will see in subsection~\ref{PixelNoiseSubsec}, the 
relevant power spectrum of the noise in channel $i$ is simply 
$C_\l^{noise,i} = 4\pi\sigma_i^2/N_i$, where 
$\sigma_i$ is the {\rms} pixel noise and $N_i$ is the number of pixels.
$C_\l^{(i)}$ denotes the power spectrum of the $i^{th}$ component at
100 GHz. In summary, the subtraction procedure is as follows:
first the maps from all frequency channels are expanded in
spherical harmonics, then the $a_{lm}$-coefficients of the various physical 
components are estimated as above, and finally the filtered maps are obtained 
by summing over these estimated coefficients, as in
\eq{MultipoleExpansionEq} with $\nu=$100 GHz.

\subsection{Pixelization and incomplete sky coverage}

All real-world CMB maps are pixelized, {\ie}, smoothed 
by some experimental beam and sampled only at a finite number of points.
In addition, the presence of ``dirty" regions such as the Galactic plane,
the Large Magellanic Cloud, 
bright point sources, {\it etc}, means that we may want to throw away
some of the pixels, leaving us with a map with a topology
reminiscent of a Swiss cheese.

In Section 5, we will see that the subtraction technique is quite efficient 
in removing the various foregrounds from a CMB map. 
The reason that it works so well is that it takes advantage of 
the fact that the foregrounds have quite different power spectra, as summarized
in \fig{EverythingFig}, by 
doing the subtraction multipole by multipole. 
With incomplete sky coverage, one cannot do quite this well, since
it is impossible to compute the exact coefficients $a_{\l m}$ 
using merely part of the sky. 
Instead of the spherical harmonics, we must 
expand our maps in some other set of basis functions, functions that vanish in
all the ``holes" in our maps. In contrast to the spherical harmonics, 
each of these
functions will inevitably probe a range of $\l$-values, rather than just a
single multipole, specified by a {\it window function} 
as described in T96.
To exploit the fact that the various foregrounds have different 
power spectra $C_\l$, we clearly want these 
window functions to be as narrow as
possible.

A prescription for how to calculate such basis functions,
taking incomplete sky coverage, pixelization, and position-dependent 
noise into account, is
given in T96, and it is found that given a patch of sky whose smallest
angular dimension is $\Delta\theta$, 
each basis function will probe an $\l$-band
of width $\Delta\l\approx 60^\circ/\Delta\theta$.
For instance, if we restrict our analysis to a 
$10^\circ\times 10^\circ$ square, then 
$\Delta\l\approx 6$. This is very good news.
It means that the only 
performance degradation in the subtraction technique will stem
from
the fact that it is unable to take advantage of sharp
features in the power spectra of width 
$\Delta\l\approx 6$ or smaller. 
This is essentially no loss at all, since 
as discussed in Section~\ref{ForegroundsSec}, we expect all the
foregrounds to have fairly smooth power spectra, without any sharp 
spikes or discontinuities. 
\ignore{
(Besides, who would trust an analysis that
was based on the assumption that we could accurately model such 
sharp features in the foreground power spectra?)
}

Whatever set of orthonormal basis functions is chosen for the analysis, 
our multi-frequency subtraction prescription
is to expand all maps in these functions and do the estimation
separately for each of the expansion coefficients.  
For any one basis function
(corresponding to a set of weights $w_k$, one for each pixel $k$), 
there will be such a coefficient $a_i$ for each of the $m$
frequency channels, and we combine these into the 
$m$-dimensional vector $\va$.
The subtraction now decomposes into the following steps:
\begin{enumerate}
\item
Compute $\sigma^2_i$, the variance in $a_i$ that is due to pixel noise
(this variance is simply a weighted sum of  
the noise variance in each pixel, the weights being $w_k^2$).

\item
Compute $\Delta_i^2$, the 100 GHz variance in $a_i$ that is due to 
the $i^{th}$ physical component, as in T96 (this variance
depends only on the power spectra and the weights $w_k$). 

\item
Compute the estimated coefficients for the $n$ different components,
denoted $\va'$.

\end{enumerate}
The last step is of course analogous to the cases we discussed above, {\ie},
$\va' = W\va$, where 
\beqa{GenWienerResultEq}
\W&=&S F^t [FSF^t + N]^{-1},\\
%G_{ij}&\equiv&\sum_{k=1}^n \F_{ik}\F_{jk}\Delta_k^2,\\
N_{ij}&\equiv&\delta_{ij}\sigma_i^2,\\
S_{ij}&\equiv&\delta_{ij}\Delta_i^2.
\label{GenWienerResultEq4}
\eeqa
Let us illustrate this with the simple example of a small square region,
sampled in a square grid of $N\times N$ points with say $N=512$. 
A convenient set of basis functions is then the discrete Fourier basis,
for which 
our subtraction would reduce to the following steps:
\begin{enumerate}
\item
Fast Fourier transform (FFT) the data.
\item
Filter as above, separately for each of the $N^2$ Fourier coefficients.
\item
Perform an inverse FFT to recover the filtered maps.
\end{enumerate}
To do still better, we can use the optimal basis functions of T96.
For this simple case, they turn out to be simply the Fourier basis
functions, but 
weighted by a two-dimensional cosine ``bell" $\cos x\cos y$ so that 
they go smoothly to zero at the boundary of the square.
Thus the prescription becomes
\begin{enumerate}
\item Multiply by cosine bell
\item FFT
\item Filter
\item Inverse FFT
\item Divide by cosine bell
\end{enumerate}
The resulting map will be very accurate in the central parts of the square,
but the noise levels will explode towards the edges where the
cosine bell goes to zero.
Thus the way to make efficient use of this technique is to tile the 
sky into a mosaic of squares with considerable overlap, so that one
can produce a low noise composite map using only the central regions of each
square.

\subsection{Non-Gaussianity and lack of translational invariance}
\label{NonGaussSec}

In the above treatment, we assumed that the statistical properties of
all random fields were translationally invariant, so that our only a
priori knowledge about them was their power spectra.  In reality, this
is of course not the case. A flagrant counterexample is the Galactic
plane, where we expect much larger fluctuations in the dust,
synchrotron and free-free components than at high Galactic latitude.
In addition, most of the foregrounds exhibit non-Gaussian behavior.
We wish to emphasize that for the purposes of estimating the
underlying CMB-fluctuations, all of these features work to our
advantage.  If we know the power spectrum of a contaminant, then
translational invariance and Gaussianity means that we have no
additional knowledge whatsoever about the contaminant, since 
the power spectrum defines the random field completely.
Clearly, the more we know about our enemy, the greater our ability
will be to tackle it and distinguish it from CMB fluctuations.

The type of non-Gaussianity that we encounter in both diffuse
components and point sources manifests itself in that the trouble is
more spatially localized than it would be for a Gaussian random field
with the same power spectrum. A bright point source affects only a
very small region of the celestial sphere, of the order of the beam
width, which can simply be removed (perhaps by using higher resolution
observations at lower sensitivities, see {\it e.g.} O'Sullivan {\etal}
1995).  Dust emission, free-free emission and synchrotron radiation
tends to be localized to ``dirty regions", with the fluctuation levels
in ``clean regions" in some cases being orders of magnitude lower.
Again, we can take advantage of this non-Gaussianity (and lack of
translational invariance when we know the cause of the emission, such
as the Galactic plane), to simply remove these regions from the data
set.  After this initial step, the analysis proceeds as described in
the previous subsection, using the power spectra that are appropriate
for the clean regions.  Thus we sift out the CMB fluctuations from the
foregrounds in a two-step process, by exploiting the fact that their
statistical properties are different both in real space and in Fourier
space:
\begin{enumerate}
\item
We place most of the weight on the clean regions in real space.
\item
We place most of the weight on the clean regions in Fourier space,
as illustrated in \fig{EverythingFig}.
\end{enumerate}

\subsection{How to avoid distorting the CMB power spectrum}
\label{RenormSec}

Although the community has displayed considerable interest in
map-making, there are of course many cases where one is merely
interested in measuring the CMB power spectrum, for instance to
constrain the parameters of theories of the formation of structure
({\it e.g.} the amplitude and spectral index of the initial
irregularities).  Rather than first generate a map with the method
presented above and then use it to estimate the power spectrum, the
latter can of course be obtained directly by aborting the reconstruction
``half way through".  Thus we can estimate $C_{\l^*}$ by first
estimating the $(2\l^*+1)$ coefficients $a_{\l m}$ that have
$\l=\l^*$, as described above, and then taking some appropriate
weighted average of the estimated $|a_{\l m}|^2$, to reduce cosmic
variance. For technical details on the choice of basis functions, the
best weights to use for the averaging, {\etc}, see T96.

When using our subtraction technique to estimate power spectra,
the normalization must be modified as described below. 
The reason for this is the above-mentioned fact that 
Wiener filtering tends to ``suck power" out of the data, 
so that the power spectrum of the
filtered map is smaller than the true power spectrum.
Moreover, this power deficit normally depends on scale,
as indicated by \eq{StandardWienerEq}.

Let us use the notation of subsection 3.2 and investigate how the 
quantity $|a'_{\l m}|^2$ is related to the power spectrum $C_\l$. 
To avoid unnecessary profusion of indices, let us focus on one single
multipole, say $\l=17$, $m=5$, and suppress the indices $\l$ and $m$ 
throughout. Thus the vector $\va$ contains the multipole 
coefficients from the $m$ different frequency channels, and 
$\va'$ the coefficients for the $n$ different physical components, as before. 
A straightforward calculation shows that (no summation implied)
\beq{PowerEstEq1}
\expec{|a'_i|^2} = |V_{ii}|^2 C^{(i)} + b_i,
\eeq
where $V\equiv WF$, and the $b_i$, the additive {\it bias}, is given by
\beq{PowerEstEq2}
b_i \equiv  
\sum_{j\neq i} |V_{ij}|^2 C^{(j)} + 
\sum_{j=1}^m   |W_{ij}|^2 C^{noise,j}.
\eeq
The power estimator 
\beq{PowerEstEq3}
C^{(i)'}\equiv |a'_i|^2 - b_i
\eeq
will thus be an unbiased estimator of the true power $C^{(i)}$, {\ie},
$\expec{C^{(i)'}} = C^{(i)}$, if we impose the normalization constraint
$V_{ii}=1$ for all $i=1,...,n$. 
As is seen in \eq{PowerEstEq2}, $b_i$ incorporates the power leakage
from the other physical components $(j\neq i)$ and from the 
pixel noise. Note that when $V_{ii}=1$, $b_i$ equals $\Delta C^{(i)}$,
the reconstruction errors of \eq{SphErrorEq}.

Let us minimize $b_i$ subject to the constraint that $V_{ii}=1$. 
Introducing the Lagrange multipliers $\lambda_i$, we thus differentiate 
$L_i \equiv b_i - \lambda_i V_{ii}$ (no summation) with respect to 
the components of the matrix $W$ and require the result to vanish. 
After a straightforward calculation, we obtain the solution 
\beq{PowerEstEq4}
W = \Lambda F^t[FSF^t+N]^{-1},
\eeq
where the matrix $\Lambda\equiv\diag\{\lambda_1,...,\lambda_n\}$. 
Imposing the normalization constraints $V_{ii}=1$ 
now gives $\lambda_i = 1/(F^t[FSF^t+N]^{-1} F)_{ii}$.
Comparing this to \eq{SphWienerResultEq}, we draw the 
following conclusion:
\begin{itemize}
\item
{\it 
Our optimized power spectrum estimate uses the same reconstruction 
matrix $W$
that we derived previously, except that the row vectors should 
be rescaled so that $(WF)_{ii} = 1$. 
}
\end{itemize}
(The extra matrix $S$ in \eq{SphWienerResultEq} is of course irrelevant
here, as it is diagonal and can be absorbed into $\Lambda$.)
This is the normalization that has been used in 
\fig{MethodsFig}. 
Since one of the main purposes of CMB sky maps is to serve as an
easy-to-visualize compliment to the 
power spectrum, we strongly advocate using the above normalization 
convention $(WF)_{ii} = 1$ when generating sky maps as well. 
As we saw above, this will ensure that CMB fluctuations in the map 
will retain their true power spectrum, rather than suffer the $\l$-dependent 
suppression characteristic of Wiener filtering.

Let us compare this with the situation in standard Wiener filtering, 
which corresponds to $m=n=1$. In this simple case, $F$ and $W$ 
are merely scalars, 
so the normalization condition gives $V=WF=1$. Thus $W$ equals a constant
$1/F$ which is independent of $\l$, corresponding 
to no subtraction at
all. In other words, if there is only one frequency channel, 
our subtraction method is of no use for power spectrum estimation. 
In the general case, there are $m\times n$ components in 
$W$ and $n$ constraints $V_{ii}=1$, so the 
subtraction method will help whenever
$m$, the number of channels, exceeds one. 

An attractive feature of the reconstruction method given by 
\eq{PowerEstEq4} is that it gives a reconstructed CMB map that
is completely independent
of our assumptions about the CMB power spectrum.
More formally, $W_{ij}$ is independent of $S_{ii}=C^{(i)}$.
This might seem surprising, since $S$ enters in the right-hand side of 
\eq{PowerEstEq4}. The easiest way to prove this result is to  
note that since the optimization problem is independent 
of the assumed CMB power spectrum (both the target function
$b_i$ and the constraint equation $(WF)_{ii}$ 
are independent of $C^{(i)}$), its solution 
(the $i^{th}$ row of $W$) must be as well.

Above we chose our filter to minimize $b_i$, the total contribution 
from the other physical components and pixel noise.
This of course produces a robust power spectrum estimator, since
if our estimate of the power spectrum of some contaminant (or our estimate of
the noise level of some channel) is off by some number of percent, 
the resulting error will scale as the corresponding term in $b_i$, 
{\eg}, 
as $|V_{ij}|^2 C^{(j)}$ (or as $|W_{ij}|^2 C^{noise,j}$). 
If one is confident that there are no such systematic errors, 
one may instead opt to minimize the 
variance of our estimator $C'_i$, which
is equivalent to minimizing the variance of $b_i$. 
This would lead to a system of cubic (rather than linear) equations
for the components of $W$, to be solved numerically.

\newpage
\section{POWER SPECTRA OF THE FOREGROUNDS}
\label{ForegroundsSec}

In this Section, we make estimates of the angular power spectra $C_\l(\nu)$
for the various foregrounds. The results are plotted in 
Figures 3 though 6, and summarized in \fig{EverythingFig}.
The former are truncated at $[\l(2\l+1)C_\l/4\pi]^{1/2}
= \sqrt{2}\times 20\muK \approx 28\>\muK$, which approximately
corresponds to COBE-normalized scale-invariant temperature fluctuations. 
Thus the
shaded regions in \fig{EverythingFig} 
are simply the top contours of Figures 3 though 6.
It should be emphasized that these estimates are {\it not} 
intended to be very accurate, especially when it comes to 
normalization. Rather, the emphasis is on their 
{\it qualitative} features, especially those that differentiate them
from one another. Despite the fact that we 
currently lack accurate high-resolution data in many important 
frequency bands, we will see that quite robust qualitative conclusions
can be drawn about which regions of the $\l-\nu$-plane will be
most suitable for estimating various parts of the CMB power spectrum.

\subsection{Point sources}

\def\nbar{{\bar{n}}}
\def\fluxcut{\phi_c}

In this Section, we make estimates of the angular power spectrum
$C_\l(\nu)$ for point sources. Here the $\l$-dependence is well known,
but the $\nu$-depencence quite uncertain. 
However, despite these uncertainties, we will see that
radio point sources will be contribute mainly to 
the lower right corner of \fig{EverythingFig}, whereas
infrared point sources will contribute mainly to the upper right.

If at some frequency there are $N$ point sources Poisson distributed 
over the whole sky, all with the same flux $\phi$, is is easy to show that 
\beq{Radio_almEq1}
\expec{a_{\l m}}=\cases{
\sqrt{4\pi}\nbar\phi&if $\l =0$,\cr
0		&if $\l\neq 0$,
}
\eeq
where $\nbar\equiv N/4\pi$ is the average number density per steradian,
and 
\beq{RadioClEq1}
C_\l \equiv \expec{|a_{\l m}|^2} - |\expec{a_{\l m}}|^2
= \nbar\phi^2.
\eeq
In other words, this would produce a simple white-noise power spectrum, 
with the same power in all multipoles, 
together with a non-zero monopole caused by the fact that no fluxes 
are negative.
If there are two independent Poisson populations, with densities
$\nbar_1$ and $\nbar_2$ and fluxes $\phi_1$ and $\phi_2$, both the means and 
the variances will of course add, giving a monopole 
$\sqrt{4\pi}(\nbar_1\phi_1 + \nbar_2\phi_2)$ and a power spectrum
$C_\l = \nbar_1\phi_1^2 + \nbar_2\phi_2^2$.
Taking the limit of infinitely many populations, 
we thus obtain 
\beqa{RadioMonopoleEq}
\expec{a_{00}}&=&
\sqrt{4\pi}\int_0^{\fluxcut}{\partial\nbar\over\partial\phi}\phi\, d\phi,\\
\label{RadioClEq2}
C_\l&=&\int_0^{\fluxcut}{\partial\nbar\over\partial\phi}\phi^2 d\phi,
\eeqa
where 
$\partial\nbar\over\partial\phi$ is the 
{\it differential source count}. In other
words,  
we have defined $\nbar(\phi)$ as the number density per steradian
of sources with flux less than $\phi$.
In real life, we are of course far from powerless against these point
sources, and can either attempt to subtract them by using spectral information
from point source catalogues, or simply choose to throw away all pixels 
containing a bright point source. In either case, the end result would be that 
we eliminate all sources with a flux exceeding some
flux cut $\fluxcut$, which then becomes the upper limit of integration
in equations\eqnum{RadioMonopoleEq} and\eqnum{RadioClEq2}.
We have estimated the source counts at $1.5\>\GHz$ from a
preliminary point source catalog from the VLA FIRST
all sky survey
(Becker {\etal} 1995). 
This catalog contains 16272 radio sources in a
narrow strip $110^{\circ} < \ra < 195^{\circ}$, 
$28.5^{\circ} < \dec < 31.0^{\circ}$, complete down to a flux limit of
$0.75\>\mJy$. A flux histogram is plotted in 
\fig{FluxFig}, together with a simple double power law fit
\beq{LumFuncFitEq}
{\partial\nbar\over\partial\phi} \approx
{524000\over\mJy\>\sr} \left(\phi\over 0.75\mJy\right)^{-1.65}
\left(1 + {\phi\over 100\mJy}\right)^{-1}
\eeq
that will be quite adequate for our purposes.
We can obviously never eliminate {\it all} radio sources, as there 
is for all practical purposes an infinite number of them,
the integral of the differential source count diverging 
at the faint end. There is also a rather obvious lower
limit to $\fluxcut$ in practice.
Since the highest resolution COBRAS/SAMBA channels have a FWHM
of $4.5$ arcminutes (see Table 1, Section 4.3.3 below), 
there are only about $10^7$ independent pixels in
the sky. Assuming that the above-mentioned FIRST data is representative of 
the entire sky, there are about 6 million sources brighter than
$0.75\>\mJy$, so if we choose $\fluxcut$ much lower than this and reject all 
data that is contaminated at this level, we would have to throw away almost
all our pixels. The subtraction strategy also has 
its limits, quite apart from the large amount of work that would 
be involved:
if we try to model and subtract the sources, it 
appears unlikely that we will be ever to do this with an accuracy exceeding
$1\%$, and even $10\%$ could prove difficult given complications such as
source variability.
Since the choice of flux cut will depend on the level of ambition of
future observing projects, 
we simply compute how the resulting power spectrum
depends on the flux cut $\phi_c$. To give a rough idea of what flux cuts
may be practically feasible in the near future, the number of radio 
sources in the entire sky are about $4\tento{6}$ above $1\,\mJy$, 
$8\tento{5}$ above $10\,\mJy$, $7\tento{4}$ above $100\,\mJy$ and
$800$ above $1\,\Jy$, all at $1.5\>\GHz$. 
The result, computed using equations\eqnum{dBdTeq},
\eqnum{RadioMonopoleEq},
\eqnum{RadioClEq2},
and\eqnum{LumFuncFitEq},
is shown in \fig{RadioClFig}.
Notice that the fluctuations have quite a different magnitude and 
$\fluxcut$-scaling than the monopole, since the two are dominated
by quite different parts of the differential source count
function. Since the slope is close to $-2$, the monopole, the total
brightness,  gets similar
contributions from several different decades of flux, whereas the 
fluctuations are strongly dominated by the brightest sources.
Thus we need not worry about not knowing the 
exact differential source count at the faint end, as all
that really matters is its behavior immediately below our 
flux cut.

Some authors ({\eg}, Franceschini {\etal} 1989) have raised the
possibility that point source clustering could create more large-scale
power that the Poisson assumption would indicate.  We have tested this
by computing the power spectrum of the FIRST data, and find no
evidence for any departure from Poisson noise (see also Benn \& Wall
1995). This conclusion, which of course simplifies the issue
considerably, is not  surprising, because most of the sources are
located at very large distances.  Correlations in the 
two-dimensional galaxy distribution that we observe (and which is the
relevant quantity when it comes to CMB contamination) are therefore
diluted by projection to negligible levels.

Although there is good reason to believe that the power spectrum 
will remain Poissonian at the higher frequencies that are relevant to the CMB, 
the issue of its normalization is of course quite complex, given the
uncertainties about the spectra and the evolution of the various 
galaxy and AGN populations (see Franceschini {\it etal} 1991).
% For a recent review of these issues, see (REF and references therein).
In \fig{pointsourcesFig}, we have simply made a $100\>\mJy$ flux cut
at $1.5\>\GHz$ (for an all-sky survey, this corresponds 
to removing about 70000 sources)
and extrapolated to higher frequencies 
with a power law $B(\nu)\propto\nu^{-\alpha}$, 
thus obtaining
\beq{radioApproxEq}
\left[{2\l+1\over 4\pi}\l\Cps_\l(\nu)\right]^{1/2} \approx
0.30\K \> \left({\sinh^2(x/2)\over 
[\nu/1.5\,\GHz]^{4+\alpha}}\right)\l,
\eeq
where $x =  h\nu/kT_0$ as before.
For $\l=100$, this corresponds to $1.5\,\muK$ at 100 GHz
if $\alpha=0$.
Lowering the flux cut to $10\>\mJy$ (removing about 900000 sources)
reduces this by about a factor 4, and 
a $1\>\mJy$ cut (removing about 5 million sources) 
gains us another factor of four.
Obviously, ambitious flux cuts become 
feasible if only a small fraction of the sky is surveyed.
Flat-spectrum sources with spectral index 
$\alpha\approx 0.3$ are likely to dominate at higher frequencies
(Franceschini {\etal} 1989), but this is of course
only to be used as a crude first approximation,
as the emission at higher frequencies is likely to be dominated by sources
whose spectra rise and peak near those frequencies,
%(Saunders 199?) 
and very little
is known about the abundances of such objects.  
We make the rather cautious assumption of an effective spectral index
$\alpha=0.0$ for the population as a whole. 
This approximation
is of course quite 
unsatisfactory at the high-frequency end, where infrared 
emission from high redshift galaxies could play an important role.
For instance, if this emission is dominated by dust in these galaxies with
emissivity $\beta=2$ (see the following Section), 
we would expect $\alpha=-4$ to be a better description 
at the higher microwave frequencies.
Unfortunately, the differential source counts of such infrared
point sources around 100 GHz is still completely unknown.
For a recent review of these issues, see Franceschini {\etal} (1991). 

\subsection{Diffuse Galactic sources}

In this Section, we discuss the qualitative features we expect for the
angular power spectra of the diffuse Galactic contaminants, namely
dust, free-free emission and synchrotron radiation.

\subsubsection{Power spectrum}

We have estimated the power spectrum of Galactic dust from 
a large number of $12.5^{\circ}\times 12.5^{\circ}$ fields of the 
100 micron IRAS all-sky survey (Neugebauer {\etal} 1984), 
which have an angular resolution
of two arcminutes (about twice as good as the best COBRAS/SAMBA 
channels). Although the amplitude varies greatly with Galactic latitude,
the overall shape is strikingly  independent of latitude,
and typically declines as $C_\l \propto 1/\l^3$ for 
$\l$ between 100 and a few thousand, steepening 
slightly on the smallest scales. This agrees well with previous 
findings (Low \& Cutri 1994; Guarini {\etal} 1995). 
To estimate the power spectrum of synchrotron radiation, 
we used the Haslam 408 GHz map (Haslam {\etal} 1982). Although the angular resolution
of this map is only of order $0.85^{\circ}$, \ie, far too low to provide
information for $\l\gg 100$, the logarithmic 
slope was found to be consistent with that 
for dust in the overlapping multipole range; around $-3$.
These results are hardly surprising, since even without analyzing 
observational data, one may
be able to guess the qualitative features of the
power spectra of the three diffuse components.
Since they are all caused by emission from diffuse blobs, one might
expect their power spectra to exhibit the following
characteristic features:
\begin{itemize}

\item $C_\l$ independent of $\l$ for small $\l$, corresponding 
to scales much greater than the coherence length of the blobs 
(this is the standard Poisson behavior, and follows if one assumes
that well separated blobs are uncorrelated).

\item $C_\l$ falls off at least as fast as $1/\l^4$ for very
large $\l$, corresponding to scales much smaller than typical
blob sizes (this follows from the simple assumption that the 
brightness is a {\it continuous} function of position).

\item If $\l^2 C_\l$ thus decreases both as $\l$ gets small and as 
$\l$ gets large, it must peak at some scale, a scale which we refer to as 
the coherence scale. 

\end{itemize}
The behavior of the contaminant power spectrum for very small $\l$ 
(whether there is indeed a coherence scale, {\it etc}), is of
course quite a subtle one, as the presence of the 
Galactic plane means that the answer will be strongly dependent on 
which patches of sky we choose to mask out during the analysis. We will return
to this issue in the subsection about non-Gaussianity below. 
In the figures, we have simply assumed that all three components have
a coherence scale of about $10^{\circ}$, corresponding to 
$\l\approx 10$, and used power spectra of the simple 
form $C_\l\propto (5+l)^{-3}$.

\subsubsection{Frequency dependence}

The frequency dependence of the three components has been 
extensively discussed in the literature
(see {\eg} Reach {\etal} 1995 and references therein).
For synchrotron radiation and free-free emission, we use simple power laws 
$B(\nu)\propto \nu^{-\beta}$. 
For synchrotron emission, $\beta\approx 0.75$ below 
10 GHz (de Bernardis {\etal} 1991), steepening to $\beta\sim 1$
above 10 GHz (Banday \& Wolfendale 1991), so we simply assume 
$\beta=1$ here.
% normalizing so that mean brightness (the monopole)
% is $0.6\mK$ in the galactic plane at 30 GHz (Wright {\etal} 1991).  
For free-free emission, we make the standard assumption $\beta=0.1$.
For dust, we assume a spectrum of the standard form 
\beq{DustSpecEq}
\Cdust_\l\propto {\nu^{3+\beta}\over e^{h\nu/kT}-1}.
\eeq
Although an emissivity index $\beta=2$ is found to be a good
fit in the Galactic plane (Wright {\etal} 1991), we use instead
the more conservative 
parameters $T = 20.7\K$, $\beta=1.36$, which are found to 
better describe the data at high Galactic latitudes (Reach {\etal} 1995),
since it is of of course the cleanest 
regions of the sky that are the most 
relevant ones for measurement of CMB fluctuations.
% Clearly, much work remains to be done, etc.

\subsubsection{Non-Gaussianity and inhomogeneity}

The spatial distributions of synchrotron radiation, free-free and dust
emission of course exhibit strong non-Gaussian features, and also a strong
departure from translational invariance because of the 
Galactic plane. 
As discussed in Section 3, this is good news  regarding our ability 
to estimate CMB fluctuations. However, it forces us to be careful when 
presenting plots of estimated power spectra. 
Thus  plots showing foreground contributions to
the CMB,  such as those presented in this paper, 
should be read with the following two caveats in mind.

First, one of the manifestations of the type of non-Gaussianity that these components
display is the presence of ``clean regions" and 
``dirty regions". For instance, a raw histogram of the brightness per 
pixel in the DIRBE 240 micron map shows that although 
$3\%$ of the pixels have a brightness exceeding 100 MJy/sr,  
the mean is only about 6 MJy/sr. The extremely bright pixels are
of course mainly located in the Galactic plane, but the level of  
``cleanness" also exhibits strong variations with Galactic longitude, caused
both by known objects such as the Large Magellanic Cloud and the
North Galactic Spur, and by the non-Gaussian clumpiness
of the dust component itself. Similar 
conclusions follow from an analysis of the IRAS
100 micron maps.
The result of this is that although the power spectrum may have a similar shape
in clean and dirty regions, the normalization will vary considerably, much
more than it would due to sample variance in a Gaussian field.
It is thus important that plots of $C_\l$-estimates are supplemented with 
a description of what type of region they refer to. In the figures in this
paper, all such power spectra refer to averages for 
{\it the cleanest 20\% of the sky}.  

Secondly, when estimating the lowest multipoles, it is important to
use as much of the celestial sphere as possible, to keep the window
functions in $\l$-space narrow (T96).  Of course, the contribution of
Galactic foregrounds increases as one includes more sky, but this is
unlikely to be a serious problem. The cleanest two thirds of the DIRBE
240 micron pixels have an average brightness about three times that of
the cleanest $20\%$ and this is a large enough area to recover all
multipoles fairly accurately except the quadrupole and octupole (T96).
Thus if we increase the sky coverage to estimate the lowest multipoles
more  accurately, the Galactic foregrounds are likely to be within a factor
of a few of the contributions from the cleanest regions of the sky,
and perhaps less if the contaminant power $\l^2 C_\l$ falls of on
scales larger than some coherence scale.

\subsection{The effects of discreteness, pixel noise and beam smoothing}

Although we usually think of pixel noise as a problem of a different
nature than the other contaminants, it can be described by an angular
power spectrum $\Cnoise_\l(\nu)$ and so be treated on an equal
footing.  One may ask what is the point of doing this, since the
statistical impact of the noise on the subtraction described in this
paper is straightforward to calculate anyway.  The answer is that it
provides better physical intuition.  Real world brightness data is of
course discretely sampled as ``pixels" rather than smooth functions
known at every point $\rh$, but as long as the sampling is
sufficiently dense (the typical pixel separation being a few times
smaller than the beamwidth), this discreteness is merely a rather
irrelevant technical detail. It enters when we do the analysis in
practice, but our results are virtually the same as if we had 
continuous sampling.

\subsubsection{Pixel noise}
\label{PixelNoiseSubsec}

\def\noise{n}

If we estimate the angular power spectrum from a sky map containing only
isotropic pixel noise, we find that all the $C_\l$-coefficients are
equal (the white noise power spectrum), 
at least down to the angular scale corresponding to the 
inter-pixel separation. This well-known result simply reflects the 
fact that the noise in the different pixels is uncorrelated.
We will now elaborate on this in slightly greater detail.

For a CMB sky map at frequency $\nu$, pixelized into $N$ pixels
pointing in the directions $\rh_i$ and
with noise $\noise_i$,
$i=1,2,...,N$, 
one typically has to a good approximation that the $n_i$
are Gaussian random variables satisfying $\expec{n_i} = 0$ and
\beq{PixelNoiseEq}
\expec{\noise_i\noise_j} = \delta_{ij}\sigma_i^2,
\eeq
for some known numbers $\sigma_i$.
We want to eliminate this discreteness from the problem, and describe 
the noise as a continuous field 
$\Bnoise(\rh,\nu)$ instead, a random field that gets added to the
actual brightness $B(\rh,\nu)$.
More specifically, for any weight function $\psi(\rh)$ that
we use in our analysis, we want the result to be
the same whether we sum over the pixels or integrate over the
field, so we require
\beq{SumIntegralEq}
{1\over N} \sum_{i=1}^N \psi(\rh_i) n_i \approx  
{1\over 4\pi} \int\psi(\rh)\Bnoise(\vr,\nu) d\Omega.
\eeq
Fortunately, this is easy to arrange: when the pixels are placed according 
to an equal-area method (as in the COBE pixelization system),
a simple choice that works is to choose $\Bnoise(\rh)$ 
to be a white noise field satisfying
\beq{WhiteNoiseEq}
\expec{\Bnoise(\rh)\Bnoise(\rh')} = \delta(\rh,\rh'){4\pi\over N}\sigma(\rh)^2,
\eeq
where $\delta$ is the angular Dirac delta function, and 
$\sigma(\rh)$ denotes the
$\sigma_i$ corresponding to the pixel position closest to $\rh$.
Since white noise by definition has no correlations on any scale, it is easy
to see why this reproduces the basic feature of the pixel noise, {\ie}, no 
correlation between neighbouring pixels. The fact that white noise fluctuates
wildly on sub-pixel scales does not invalidate
\eq{SumIntegralEq}, since any weighting function $\psi$ that we
use in practice cannot vary on sub-pixels scales
(since we will after all apply it to pixels), 
and thus smoothes out this substructure.

For the purposes of analyzing future experiments, let us assume that
the pixel noise is independent of position, so that $\sigma_i$ 
simply equals some constant, $\sigma$. 
This means that the white noise is isotropic, and has a well-defined angular power
spectrum $\Cnoise_\l$ that we will now compute.
For white noise,  the power spectrum is 
independent of $\l$, and 
thus all we need to do is find the overall normalization, expressed in terms of
the pixel noise and number of pixels.
We do this by
examining the simplest multipole, $\l=0$.
Since $Y_{00} = 1/\sqrt{4\pi}$, \eq{MultipoleExpansionEq}
gives 
\beq{NoiseEq1}
\anoise_{00} = \sqrt{4\pi}\left({1\over 4\pi}\int \dTnoise(\rh) d\Omega\right),
\eeq
where we have factored out $\sqrt{4\pi}$ so that the expression in parenthesis
is the sky-averaged temperature. 
Replacing this by the 
pixel-averaged temperature (simply using \eq{SumIntegralEq} with $\psi=1$), we obtain
\beq{NoiseEq2}
\anoise_{00} = \sqrt{4\pi}\left({1\over N}\sum_{i=1}^N \dTnoise(\rh_i)\right).
\eeq
Using 
\eq{PixelNoiseEq},
this leaves us with our desired result,
\beq{NoiseEq4}
\Cnoise_\l = \Cnoise_0 = \bigexpec{\left|a_{00}\right|^2} = 
{4\pi\over N}\sigma^2.
\eeq
It is straightforward to verify that this normalization
agrees with that in \eq{WhiteNoiseEq}.

\subsubsection{Beam smoothing}
\def\Bobs{B^{obs}}
\def\beamf{w}
\def\beamh{\widehat{\beamf}}
\def\beamsig{\theta_b}

In this subsection, we point out that the effects of beam smoothing
can be completely absorbed into the description of the noise.

In a single-beam experiment, the 
brightness field $\Bobs$ is the true field $B$ convolved with a 
beam function $\beamf$,
\beq{BeamConvEq}
\Bobs(\rh) = \int\beamf(\theta) B(\rh') d\Omega',
\eeq
where $\theta\equiv\cos^{-1}(\rh\cdot\rh')$ is the angle between the
two vectors.
As long as the beam width is much less than a radian,
expanding this in spherical harmonics gives the familiar result
\beq{FourierBeamEq}
C_{\l}^{obs} \approx \left|\beamh(\l)\right|^2C_\l,
\eeq
where $\beamh$ is the Fourier transform of $\beamf$.
In the common approximation that the beam profile is a Gaussian,
\beq{GaussianBeamConvEq}
\beamf(\theta) = {1\over\sqrt{2\pi}\beamsig}
e^{-{1\over 2}{\theta^2\over\beamsig^2}},
\eeq
\eq{FourierBeamEq} reduces to 
\beq{FourierBeamEq2}
C_{\l}^{obs} \approx e^{-\beamsig^2\l(\l+1)} C_\l.
\eeq
The standard way to quote the beam width is to give
the full-width-half-max (FWHM) width of $\beamf$, 
so the correspondence is 
$\beamsig = \FWHM/\sqrt{8\ln 2} \approx 0.425\>\FWHM.$

\noindent
This suppression of high multipoles by beam smoothing 
of course affects the fields $B_i$ for all the different components 
($\Bcmb$, $\Bdust$, {\etc}) except one. The exception
is $\Bnoise$, since the pixel noise gets added after the
beam smoothing, and thus has nothing to do with the beam width.
Although it is easy to convert between actual and observed 
fields (using \eq{FourierBeamEq} for the power spectrum or
a simple deconvolution for the fields $B_i(\rh)$), it is quite 
a nuisance to always have to distinguish between the two.
Since $\Bnoise$ is the only field that is simpler to describe 
``as observed", we adopt the following convention:
\begin{itemize}
\item
We let all fields $B_i(\rh,\nu)$ refer to the {\it actual} fields,
as they would appear if there were no beam smoothing.
\end{itemize}
We thus define the ``unsmoothed" noise field as
\beq{UnsmoothedNoiseEq}
\Cnoise_{\l}(\nu) \equiv 
{4\pi\sigma(\nu)^2/N \over \left|\beamh\left(\l\right)\right|^2}.
\eeq
This is what is plotted in \fig{NoiseFig}.
In other words, the advantages of this convention are that
\begin{itemize}
\item
this figure can be directly compared with those for the 
various foregrounds, and
\item 
the latter figures are independent of any assumptions about
the beam width.
\end{itemize}

\subsubsection{COBRAS/SAMBA specifics}

The proposed COBRAS/SAMBA satellite mission (Mandolesi {\etal} 1995) is
currently being evaluated by the European Space Agency, 
and if approved, is scheduled for launch in 2003.
As currently proposed, it would 
have nine frequency
channels, as summarized  in Table 1. 
\begin{table}
\begin{tabular}{|l|ccccccccc|}
\hline
Channel			 &1&2&3&4&5&6&7&8&9\\
\hline
Center frequency $\nu$ [GHz] &31.5&53&90&125&143&217&353&545&857\\
Bandwidth $\Delta\nu/\nu$ &0.15&0.15&0.15&0.15&0.35&0.35&0.35&0.35&0.35\\
FWHM beam size [arcmin]&30&18&12&10&10.5&7.5&4.5&4.5&4.5\\
Pixel noise $\sigma(\nu)$, 2 years [$\mu$K] 
	&17.7&18.3&23.7&69.5&5.7&6.0&36.0&271&62700\\
\hline
\end{tabular}
\caption{COBRAS/SAMBA channel specification}
\label{ytable1}
\end{table}
Although we will use the exact 
numbers from the this table in the analysis in 
Section~\ref{ResultsSec}, we have made a few 
simplifying approximations in  generating 
\fig{NoiseFig}, since we want this plot to be approximately applicable
to any satellite mission using similar technology.

\vskip 0.2 truein

\noindent
{\bf Sensitivity:}
The reason that $\sigma(\nu)$ explodes for large $\nu$ is of course 
the exponential fall-off of the Planck-spectrum.
Converting $\sigma$ to brightness fluctuations using 
\eq{dBdTeq}, one finds that to a crude approximation,
$\sigma(\nu)\propto\nu$.
%\beq{sigmaAppoxEq}
%\sigma(\nu) \approx  {250\,\Jy/\sr\> (\nu/100\GHz)\over
%\partial B_0/\partial T_0}.
%\eeq
The main deviation from this power law
occurs around 130 GHz, where the 
the technology transition
from HEMT to bolometer detectors between channels 4 and 5 can be seen to
cause the noise to drop by an order of magnitude in Table 1.
In generating figures~\ref{NoiseFig} and~\ref{EverythingFig}, 
we have simply interpolated the tabulated values with a cubic spline.

\vskip 0.2 truein

\def\osamp{\eta}

\noindent
{\bf Number of pixels:}
Typically, the number of pixels is chosen so that neighbouring pixels overlap
slightly. Since the combined area of all pixels
is proportional to  $\FWHM^2\>N$,
it is therefore convenient to define the dimensionless quantity
$\osamp \equiv \FWHM^2 N/4\pi$.
In terms of $\osamp$, which is thus a measure of the rate of 
oversampling, we can write the numerator 
of \eq{UnsmoothedNoiseEq} as
\beq{OversampEq}
{4\pi\sigma^2\over N} = {\FWHM^2\sigma^2\over\eta},
\eeq
{\eg}, in terms of the quantities that are usually quoted 
in experimental specifications. 
The COBE DMR experiment had $\FWHM = 7.08^{\circ}$ 
and N=6144 pixels, which gives an oversampling factor 
$\eta\approx 7.47$.
In this paper, we will assume that the COBRAS/SAMBA experiment
will use this same degree of
oversampling, in all channels.

\vskip 0.2 truein

 \noindent
{\bf Beam width:} From Table 1, we see that the resolution is
diffraction limited (FWHM$ \propto 1/\nu$) at low frequencies,
corresponding to a mirror size of order 1 meter, 
whereas it is constant at the highest frequencies.
In generating figures~\ref{NoiseFig} and~\ref{EverythingFig}, 
we have simply interpolated the tabulated values with a cubic spline.

\ignore{
Here the simple fit 
\beq{ResFitEq}
{\beamsig\over 0.6} \approx 4'\left[1 + \left({\nu\over
140\>\GHz}\right)^{-1.3}\right]
\eeq
for the frequency dependence is adequate for the purposes of 
\fig{NoiseFig}.
Combining all these estimates, we thus obtain
\beqa{NoiseApproxEq}
\nonumber
\left[{2\l+1\over 4\pi}\Cnoise_\l(\nu)\right]^{1/2}&=&
\sqrt{2\l+1\over N} 
\exp\left[{1\over 2}\beamsig(\nu)^2\left(\l+{1\over 2}\right)^2\right] \sigma(\nu)
{\partial B_0\over\partial T_0}\\
&\approx&
0.010 \muK \> \sqrt{2\l+1} e^{\beamsig(\nu)^2\l^2/2}{(\sinh x/2)^2\over x^{2.7}},
\eeqa
where $\beamsig(\nu)$ is given by 
\eq{ResFitEq} and $x =  h\nu/kT_0$ as before.
This function is plotted in \fig{NoiseFig}.
(These fits are used for illustration only --- in the
calculations presented in the following Section, 
the exact figures from Table 1 have been used.)
}

\clearpage
\section{RESULTS}
\label{ResultsSec}

We have computed the signal-to-noise ratio obtainable with the 
technique presented in Section~\ref{WienerSec} assuming that the foregrounds
behave as conjectured in Section~\ref{ForegroundsSec}. 
We have taken $m=9$ and $n=5$, corresponding to the 
nine COBRAS/SAMBA channels specified in Table 1 and the five 
components CMB, dust, radio point sources, synchrotron radiation
and free-free emission. 

\subsection{Multi-frequency subtraction versus no subtraction}

The resulting reconstruction errors $\Delta C_l$ computed from
\eq{SphErrorEq} are shown in \fig{MethodsFig} (bottom heavy line,
delimiting the double-hatched region), and are seen to lie about three
orders of magnitude beneath a CDM power spectrum, corresponding to a
signal-to-noise ratio of about $10^3$ for CMB map reconstruction
(referring to all foregrounds collectively as ``noise" in this
context).  For estimating the power spectrum $C_\l$, a quadratic
quantity, the corresponding signal-to-noise ratio would be an
outrageous $10^6$ for $\l\simlt 1000$.  Clearly, this sort of accuracy
will not be attainable in practice, because we do not know the
frequency dependence of the various contaminants accurately enough to
trust the extrapolations involved in the subtraction.  We will return
to this issue below.  Thus interpreting this as an upper limit to how
well we can hope to do, we also wish to place a lower limit on how
poorly we can do.  The most simplistic approach possible is of course
making no attempts whatsoever at foreground subtraction, and using
merely a single channel.  The resulting errors for the 143 GHz channel
are given by the uppermost curve in the figure. A slight improvement
would be the solid curve below it, which corresponds to choosing the
single best COBRAS/SAMBA channel separately for each multipole. The
frequency where the combined foreground contribution is minimized is
plotted as a function of $\l$ in \fig{EverythingFig} (heavy dashed
line).  The reason for its positive slope is of course that when $\l$
increases, the power spectrum of dust decreases whereas the power
spectrum of radio point sources, attacking from below, increases.
This same effect is illustrated in \fig{MultipolesFig}, which shows
the total contribution from all foregrounds to the observed $C_l$ as a
function of frequency.  It is seen that although the most promising
frequency for estimating low multipoles is $\nu\sim 50$GHz, the
central COBE channel, the optimal frequency for searching for CDM
Doppler peaks ($\ell \simgt 200$) is around 100 GHz.  The situation at
three of the COBRAS/SAMBA frequencies is summarized in
Figures~\ref{Channel31Fig} through~\ref{Channel217Fig}.

\subsection{Why pixel-by-pixel subtraction performs so much worse}

An alternative subtraction method would 
be to combine the smaller pixels of the high 
channels into 30 arcminute pixels, and then simply apply our subtraction 
scheme on a pixel-by-pixel basis, using 
equations\eqnum{GenWienerResultEq}-(\ref{GenWienerResultEq4})
as follows:
\begin{enumerate}
\item 
Compute the pixel variance $\Delta_i^2$ contributed by each component,
as $\Delta_i^2 = \sum_\l (2\l+1)C_l^{(i)}|\beamh(\l+1/2)|^2/4\pi$. 
\item
Use the pixel variance of Table 1 for $\sigma_i^2$ in the matrix $N$.
\item
Use the resulting matrix $W$ to reconstruct the CMB temperature for each pixel.
\item 
Estimate $C_\l$ by expanding the resulting CMB map in spherical harmonics.
\end{enumerate}
Loosely speaking, this method corresponds to subtracting before Fourier
transforming, rather than vice versa as in the optimized method.
We found that this method produced a signal-to-noise ratio 
of about 15 for each pixel. The 
resulting reconstruction errors are shown by the heavy dashed line in
\fig{MethodsFig}, and are seen to be more than an order of magnitude worse 
than the optimized method for the small $\l$-values where the low
($30^\prime$) resolution is not a problem. 
This degradation in performance may seem surprising, since the 
same (overly optimistic) assumption that we know the frequency 
dependence of all components is made with this method. 
However, the cause is exactly the same as that of the 
discrepancy between the two
uppermost curves in the figure: the optimal channel weighting is different
for high and low multipoles. Thus the pixel-by-pixel approach selects one
single channel weighting that does not perform unduly badly for any multipole, 
at the price of not
doing especially well at any multipole either.
The optimized method, on the other hand, tailors the channel weighting separately
for each multipole.

\subsection{Why ``direct subtraction" does so poorly}

Another natural method would be to simply
select as many channels as there are components, and
reconstruct the true fields by merely inverting the matrix $F$,
{\ie}, by choosing $\vx' = F^{-1}\vy$.
Although this gives exactly the right answer in the absence of pixel
noise, this method performs very poorly when the pixel 
noise is non-negligible. The noise in the resulting
reconstruction will typically be dominated by the most noisy of all 
the channels, and when the matrix $F$ is poorly conditioned (such as when two
components like synchrotron radiation and free-free emission 
have similar spectral indices), there 
will be additional noise amplification.
If the number of frequencies $m$ equals the number of components $n$, 
then in the limit of no noise ($N=0$),  
our general subtraction scheme reduces to
\beq{NoNoiseEq}
W = SF^t(FSF^t)^{-1} = F^{-1},
\eeq
{\ie}, to simple
``direct subtraction", independently of our assumptions about
the various power levels (which are contained in $S$).  
Since for this special case, $WF=F^{-1}F=I$, we see that 
the power-preserving subtraction we advocate also reduces to
direct subtraction when the noise is ignored, since the 
normalization condition $(WF)_{ii}=1$ is satisfied.
Since the optimized method is computationally trivial anyway, involving 
merely the inversion of a small matrix,
there appears to be no advantage in neglecting the noise
and using ``direct subtraction". 

It should also be emphasized that whereas ``direct subtraction"
and least-squares generalizations thereof 
require at least as many channels as components, our more general 
method has no such restriction, allowing 
$m<n$, $m=n$ and $m>n$.

\subsection{Modeling spectral uncertainties}

The main caveat to bear in mind when using any of the above-mentioned
subtraction schemes is that, in reality, we do not know the matrix $F$
with perfect accuracy.  For instance, if the spectral index of
free-free emission is $\beta=0.2$ rather than $\beta=0.1$, a
subtraction attempt based on an extrapolation from 31 GHz to 217 GHz
will be off by about $20\%$.  We obviously expect the spectral indices
to vary slightly from one sky region to another. For example,
unsubtracted radio sources will not all have the same spectra and so their
average spectrum will vary with position.

For us to be able to trust the error bars produced by a foreground
removal method, we clearly need a way of quantifying the impact of
such uncertainties in spectral indices.  A simple way to do this is of
course to assume some probability distributions for the spectral
indices, make Monte-Carlo realizations, and compute the average of the
actual reconstruction errors obtained when (erroneously) assuming the
the spectral indices are exactly known.  However, this is merely a
method of diagnostics, and would not in itself provide a more robust
subtraction scheme.  Fortunately, there is quite a simple way of
incorporating such uncertainties into the foreground model
itself, which we will now describe.

\ignore{
The lowermost curve in \fig{MethodsFig} show the reconstruction errors
that would result if there were no synchrotron radiation or free-free
emission. The curve above this one shows the result if merely
free-free emission is absent. Notice that for very low $\l$, where we
have assumed that the two are of comparable importance at 31 GHz,
neglecting one of the components helps almost as much as neglecting
both of them.  This is because here the algorithm fails to break the
degeneracy between the two ($\beta=0.1$ versus $\beta=1.0$), faced
with noise and confusion from the true CMB fluctuations.  We can take
advantage of this very effect to incorporate our uncertainties about
spectral indices into our model.} Suppose we have reason to believe
that about half of the synchrotron emission is characterized by
$\beta\approx 1.1$ and half by $\beta\approx 0.7$.  We can simply
incorporate this into our model as two separate components with the
same power spectrum $C_\l$ but with different spectral dependencies
$f_i(\nu)$.  More realistically, we may wish to include an allowed
range of spectral indices, reflecting either our lack of knowledge of
their precise values or the presence of several physically distinct
sub-components. In either case, the way to incorporate these ranges
into the analysis is the same.  If we wish to put into the model that
$\beta = 0.9\pm 0.2$, for instance, we can simply insert two
synchrotron components, one with $\beta=0.7$ and one with $\beta=1.1$
(or a  range of components, if one prefers) and the
multi-frequency subtraction formalism will automatically reflect
our uncertainty in $\beta$ by associating appropriately large
reconstruction errors with synchrotron subtraction.

\subsection{Satellite specifics}
  
We conclude this Section with a few more technical results relating to
the impact of COBRAS/SAMBA specifics on the reconstruction errors, the
lowermost heavy curve in \fig{MethodsFig}.

Removing channel 9 makes almost no difference. Removing the three
highest frequency channels produces the thin solid curve in the same figure,
{\ie}, a worsening by a factor of a few over the whole range of
multipoles.  Removing the all the HEMT channels (channels 1-4), yields
the thin dashed line, and most of this loss of accuracy
is incurred even
if only channel 1 is removed --- basically because this greatly
reduces the ability to subtract out synchrotron radiation from the
higher frequency channels.  When all channels are included, the multi-frequency
subtraction scheme is found to place most of the weight on channels 4, 5
and 6. Although the other channels receive considerably smaller
weights, they still help considerably, as these weights are tuned so
as to subtract out the residual foregrounds.

It should be emphasized that although removing a few channels as described
above made a relatively minor difference 
when the spectral indices were assumed to be perfectly known, 
broad spectral coverage is of course of paramount importance 
to ensure that the removal process is robust and can handle
uncertainties in spectral slopes as described in the previous subsection.

%\clearpage
%%%%%%%%%%%%%%%%%%%%%% CONCLUSIONS: %%%%%%%%%%%%%%%%%%%%%%%%%

\section{DISCUSSION}

We have presented a new method for subtracting foreground
contamination from multi-frequency microwave sky maps, which can be
used to produce accurate CMB maps and estimates of the CMB power
spectrum $C_\l$. 

\subsection{Relation to other subtraction methods}

The method incorporates the noise levels and 
subtracts the
foregrounds in the Fourier, or multipole, domain rather than in real
space, thereby exploiting the fact that contaminants such as dust,
point sources, {\etc}, tend to have power spectra that differ
substantially from that of the CMB.
Compared to a standard subtraction procedure, it thus
improves the situation in two ways:
\begin{enumerate}
\item The noise levels of the input maps are taken into account.
\begin{itemize}
\item 
In a simple subtraction scheme, the 
cleaned map is just a particular linear combination of the 
input maps at various frequencies, where the weights
take no account of the noise levels. For instance, 
if both the IRAS 100 micron map and the DIRBE 140 micron map
were used as dust templates, the relative weight assigned to 
the two would be arbitrary in a simple subtraction
scheme but optized in the approach presented here.   
\item 
Simple subtraction is merely a special case of our
method applied on a pixel-by-pixel basis, obtained in the
limit of zero noise when $m=n$.
\item 
Since any spatial template whatsoever can be included
as a ``channel", the special case of simple subtraction thus 
has no advantages whatsoever over the more 
general method presented here.
\end{itemize}
\item The subtraction is performed in Fourier space.
\begin{itemize}
\item Since no phase information is lost by
passing to the Fourier (multipole) domain and back,
this preserves all the information about the various spatial 
templates, for instance the location of the Galactic plane.
\item Thus this approach is superior to working in real space 
whenever the contaminants have power spectra that differ
substantially from that of the CMB.
\item This will be the case for {\it regardless} of what the
true CMB power spectrum turns out to be, since certain contaminants are 
known to have power spectra that differ from one another
(for instance, the dust power scales roughly as $\l^{-3}$, 
whereas the point source power scales as $\l^0$.
\item Although normal Wiener-filtering modifies the 
power spectrum of the measured signal, our 
prescription in \eq{PowerEstEq4} 
(where the row vectors of $W$ are rescaled 
so that $(WF)_{ii} = 1$)
does {\it not} not have 
this undesirable property, which means that our subtraction
scheme is suitable for power spectrum estimation as
well as map-making (we showed that this is only possible when 
more than one frequency is available). 
\item
With this prescription, the optimal weighting 
coefficients in the $W$-matrix are independent of any assumptions
about the CMB power spectrum (in stark contrast 
to conventional one-channel Wiener filtering). 
The weights only depend on the assumptions
about the foreground power spectra.  
\item 
It is of course advantageous 
to perform the subtraction in Fourier space even  
if one does not wish to make any assumptions about
the foreground levels. This model-independent special case of our
method corresponds to simply setting
the noise levels to zero.
\end{itemize}

\end{enumerate}
Any of these two improvements can of course be implemented without the
other. 
Traditional subtraction, but mode by mode in Fourier space, will perform better
than if done pixel-by-pixel in real space. 
Likewise, if the overall foreground levels are known in many channels, 
the result will be better if noise is taken into 
account than if it is ignored in the subtraction.

\subsection{Foreground estimates}

To provide a qualitative understanding of how well the method will be able to
tackle the next generation of CMB data, we made rough estimates of the power
spectra of the relevant foregrounds in Section~\ref{ForegroundsSec}. 
The results are summarized in \fig{EverythingFig}.
Although these estimates are not intended to be 
very accurate, the following qualitative
conclusions appear to be quite robust:
\begin{itemize}
\item 
Galactic dust poses a problem mainly at the upper left, corresponding to 
large scales and high frequencies.
\item 
Synchrotron radiation and free-free emission pose a problem mainly at the lower
left, corresponding to large scales and low frequencies.
\item 
Radio point sources are a problem mainly at the lower right, 
corresponding to small scales and low frequencies.
\item 
If infrared emission from high redshift
 galaxies pose a significant problem, it will
be at the upper right, 
corresponding to small scales and high frequencies.
\item
Experimental noise and beam dilution are mainly a limitation to the right and 
above.
\item
The most favorable frequencies for measuring high multipoles such as the CDM
Doppler peaks are larger (around 100 GHz and above) than the best ones for
probing the largest scales (around 50 GHz). 
\end{itemize}

\subsection{An example: COBRAS/SAMBA}

In Section 5, we assessed the effectiveness of the  
method using the specifications of the proposed COBRAS/SAMBA satellite 
mission and the above-mentioned foreground modeling.
As is seen in \fig{MethodsFig}, our
method provides a gain of more than a factor of ten over the entire
multipole range compared to 
subtracting the backgrounds on a pixel-by-pixel basis.
If the frequency dependencies of the foregrounds were perfectly
known and independent of position in the sky (which they are not), 
then the method
could recover the CMB multipoles $a_{\l m}$ out to 
about $\l=10^3$ to an accuracy of about a tenth of a percent.
We also found that our uncertainty about the 
spectral indices could be incorporated
into the formalism in straightforward way, by simply replacing a poorly
understood component
by several components, all with the same power spectra, but with 
slightly different spectral indices. 
We saw that even with extremely pessimistic assumptions about our 
ability to model the
foregrounds, it is likely that an estimate of the CMB power
spectrum $C_\l$
can be made where the residual foreground contamination is merely a few
percent.
Since 
the power spectrum of dust falls off and that of point 
sources rises
with $\l$ (relative to that
of the CMB), the most favorable situation occurs around 
$\l=200$, where even power 
accuracies of $1\%$ would be an extremely pessimistic
prediction.
This is fortunate, as this scale coincides with the location of 
the first Doppler peak of standard CDM, and accurate determination of its
position (if some variant of CDM is correct) 
would provide a direct measurement
of $\Omega$, the cosmological density parameter
(see {\it e.g.} Kamionkowski \& Spergel 1994).

\subsection{The effect of non-Gaussianity}

The case where the foregrounds are Gaussian is in a sense the worst 
possible case when it comes to subtracting them.
Since Gaussian random fields are completely specified by their power spectra, 
this means that we have no additional information that we can take advantage
of in our attempts at foreground removal.
Since the method we derived was optimized for the Gaussian case, 
a natural question to ask is whether one can do still better by making use
of the fact that many foregrounds do in fact exhibit non-Gaussian features.
One simple way to do this is to simply remove spatially localized
contaminants (\eg, the Galactic plane and bright point sources),
as was discussed in Section~\ref{NonGaussSec}. 
Making optimal use of non-Gaussian features, however, 
is quite difficult in practice, as it generally leads to a 
non-linear optimization problem. Even if a numerical solution
were found, one might be left wondering to what extent 
poor assumptions about 
the precise type of non-Gaussianity might have degraded the final results.

Fortunately, the method we have presented appears to be quite
adequate in practice. As part of the scientific
case for the COBRAS/SAMBA satellite, detailed simulations have been made 
where real foreground maps (extrapolated to the relevant frequencies
using the data from IRAS, the Haslam map, {\etc}) were added to 
simulated CMB data. The simulated real-world maps were then 
``cleaned" with various subtraction 
methods. The results (Bouchet {\etal} 1996) show that
our method works extremely well, even though the IRAS dust maps 
are known to exhibit strong
non-Gaussian features.

\subsection{Outlook: what more do we need to know?}

To be able to better quantify the effectiveness of future CMB missions
and answer questions such as ``what design changes would 
improve the ability to remove foregrounds the most?", it is important that 
more accurate measurements of the various foreground power spectra 
be made.
Experiments are now becoming so sensitive that it no longer suffices to
summarize each foreground by a single number $\Delta T$ 
(which usually refers to its average intensity, the monopole) 
--- its entire power spectrum is needed, to chart out the 
foreground landscape
of \fig{EverythingFig}.
Some of the most urgent outstanding questions are the following:
\begin{itemize}
\item
What is the differential source count of radio point sources between 
10 GHz and 100 GHz, {\ie}, how do we normalize their power spectrum 
for various flux cuts?
\item 
What is the differential source count 
of infrared point sources between 50 and 300 
GHz, {\ie}, how do we normalize their power spectrum 
for various flux cuts?
\item
What is the power spectrum of synchrotron radiation and free-free emission on
small angular scales?
\end{itemize}
If none of the foreground contaminants turn out to be much worse 
than we assumed in Section 4,  
then the next generation of CMB experiments may indeed allow us to measure
the key cosmological parameters with hitherto unprecedented accuracy.

\vskip 0.5 truein

\noindent
{\bf ACKNOWLEDGMENTS:} 
We thank Ted Bunn, Fran\c{c}ois Bouchet, 
Anthony Lasenby, Jean-Lup Puget and Ned Wright 
for many useful comments and 
Gavin Dalton for his help
with the IRAS $100\mu$m and DIRBE maps. 
MT acknowledges partial support from European Union contract
CHRX-CT93-0120 and Deutsche Forschungsgemeinschaft
grant SFB-375.

%\clearpage
%%%%%%%%%%%%%%%%%%%%%% REFERENCES: %%%%%%%%%%%%%%%%%%%%%%%%%

\section{REFERENCES}

\rf Banday, A. J. \& Wolfendale, A. W. 1991;MNRAS;248;705

\rf Becker, R. H., White, R. L. \& Helfand, D. J. 1995;ApJ;450;559
% The FIRST Survey: Faint Images of the Radio Sky at 20 cm
% Robert H Becker, Richard L White & David J Helfand
% ASTROPHYSICAL JOURNAL, 1995 SEP 10, V450 N2:559-577.

\rf Benn, C.R. \& Wall, J.V. 1995;MNRAS;272;481

\rf Bond, J.R. \& Efstathiou, G., 1987;MNRAS;226;655

% \rf Bond, J. R. {\etal} 1994;Phys. Rev. Lett.;72;13

\rf Bond, J. R. 1995;Phys. Rev. Lett.;74;4369
% Signal-to-Noise Eigenmode Analysis of the Two-Year COBE Maps

\rf Bouchet, F. {\etal} 1995;Space Science Rev.;74;37
% preprint astro-ph/9410004
% SIMULATIONS OF THE MICROWAVE SKY AND OF ITS ``OBSERVATIONS'' by
% F.R. Bouchet, R. Gispert, N. Aghanim, J.R. Bond, A. De Luca, E. Hivon, and B. Maffei, 6
% pages of uuencoded compressed postscript (1.2 Mb uncompressed), to appear in the
% proceedings of the meeting "Far Infrared and Sub-millimeter Space Missions in the Next
% Decade'', Paris, France, Eds. M. Sauvage, Space Science Review, 
% BOUCHET FR; GISPERT R; AGHANIM N; BOND JR; and others.
% SIMULATIONS OF THE MICROWAVE SKY AND OF ITS OBSERVATIONS.
% SPACE SCIENCE REVIEWS, 1995 OCT, V74 N1-2:37-43.
% Pub type:  Review.
    
\rn Bouchet, F. {\etal} 1996, in preparation.
% SIMULATIONS OF THE MICROWAVE SKY AND OF ITS ``OBSERVATIONS'' by
% F.R. Bouchet, R. Gispert, N. Aghanim, J.R. Bond, A. De Luca, E. Hivon, and B. Maffei, 6
% pages of uuencoded compressed postscript (1.2 Mb uncompressed), to appear in the
% proceedings of the meeting "Far Infrared and Sub-millimeter Space Missions in the Next
% Decade'', Paris, France, Eds. M. Sauvage, Space Science Review, 

\rf Brandt, W. N. {\etal} 1994;ApJ;424;1 
% Brandt, W.N.; Lawrence, C.R.; Readhead, A.C.S.; Pakianathan, J.N.; and others.
% Separation of foreground radiation from cosmic microwave background
% anisotropy using multifrequency measurements.
% Astrophysical Journal, 20 March 1994, vol.424, (no.1, pt.1):1-21.

\rf Bunn, E. F. {\etal} 1994;ApJ;432;L75
% Bunn, E.F.; Fisher, K.B.; Hoffman, Y.; Lahav, O.; and others.
% Wiener filtering of the COBE differential microwave radiometer data.
% Astrophysical Journal, Letters, 10 Sept. 1994, vol.432, (no.2, pt.2):L75-8.
     
\rf Bunn, E. F., Scott, D. \& White, M. 1995;ApJ;441;L9
% THE COBE NORMALIZATION FOR STANDARD CDM, CfPA-94-TH-42 

\rf Bunn, E. F. \& Sugiyama, N. 1995;ApJ;446;49
% preprint astro-ph/9407069
% Lambda
% COSMOLOGICAL CONSTANT COLD DARK MATTER MODELS AND THE COBE TWO-YEAR SKY MAPS.
% ASTROPHYSICAL JOURNAL, 1995 JUN 10, V446 N1:49-53.

\rf de Bernardis, P., Masi, S. \& Vittorio, N. 1991;ApJ;382;515

\rf Dodelson, S. \& Stebbins, A. 1994;ApJ;433;440
% Analysis of small-scale microwave background radiation anisotropy in the
% presence of foreground contamination.
% Astrophysical Journal, 1 Oct. 1994, vol.433, (no.2, pt.1):440-53.

\rf Fisher, K. B. {\etal} 1995;MNRAS;272;885
% Fisher, K.B.; Lahav, O.; Hoffman, Y.; Lynden-Bell, D.; and others.
% Wiener reconstruction of density, velocity and potential fields from
% all-sky galaxy redshift surveys.
% Monthly Notices of the Royal Astronomical Society, 
% 15 Feb. 1995, vol.272, (no.4):885-908.

\rf Franceschini, A. {\etal} 1989;ApJ;344;35
% FRANCESCHINI A; TOFFOLATTI L; DANESE L; DEZOTTI G.
% DISCRETE SOURCE CONTRIBUTIONS TO SMALL-SCALE ANISOTROPIES OF THE
% MICROWAVE BACKGROUND.
% ASTROPHYSICAL JOURNAL, 1989 SEP 1, V344 N1:35-45.
      
\rf Franceschini, A. {\etal} 1991;A\&A Supp.;89;285
% Franceschini, A.; Toffolatti, L.; Mazzei, P.; Danese, L.; and others.
% Galaxy counts and contributions to the background radiation from 1 mu m to
% 1000 mu m.
% Astronomy & Astrophysics Supplement Series, Aug. 1991, vol.89, (no.2):285-310.
     
\rf G\'orski, K. M. 1994;ApJ;430;L85
%  On determining the spectrum of primordial inhomogeneity from the COBE DMR
%  sky maps: method.     

\rf Guarini, G., Melchiorri, B. \& Melchiorri, F. 1995;ApJ;442;23
% GUARINI G; MELCHIORRI B; MELCHIORRI F.
% INFLUENCE OF GALACTIC DUST ON THE ACCURACY OF COSMIC MICROWAVE ANISOTROPY
% MEASUREMENTS.
% ASTROPHYSICAL JOURNAL, 1995 MAR 20, V442 N1:23-29.
     
\rf Haslam, C. G. T. {\etal} 1982;A\&A Supp.;47;1 % Check the initials!      
     
\rf Hu, W. \& Sugiyama, N. 1995;ApJ;436;456
% preprint astro-ph/9411008
% Toward understanding CBR anisotropies and their implications

\rf Jungman, G, Kamionkowski, M., Kosowsky, A \& 
Spergel, D.N., 1996;Phys. Rev. Lett.;76;1007
% JUNGMAN G; KAMIONKOWSKI M; KOSOWSKY A; SPERGEL DN.
% WEIGHING THE UNIVERSE WITH THE COSMIC MICROWAVE BACKGROUND.
% PHYSICAL REVIEW LETTERS, 1996 FEB 12, V76 N7:1007-1010.     
     
\rf Kamionkowski, M. \& Spergel, D.N., 1994;ApJ;431;1

\rf Lahav, O. {\etal} 1994;ApJ;423;L93
% Lahav, O.; Fisher, K.B.; Hoffman, Y.; Scharf, C.A.; and others.
% Wiener reconstruction of all-sky galaxy surveys in spherical harmonics.
% Astrophysical Journal, Letters, 10 March 1994, vol.423, (no.2, pt.2):L93-6.

%\rf Lineweaver, C. H. {\etal} 1994;ApJ;436;452
% LINEWEAVER CH; SMOOT GF; BENNETT CL; WRIGHT EL; and others.
% CORRELATED NOISE IN THE COBE DMR SKY MAPS.
% ASTROPHYSICAL JOURNAL, 1994 DEC 1, V436 N2:452-455.

\rf Low, F. J. \& Cutri, R. M. 1994;Infrared Phys. \& Technol.;35;291

\rf Mandolesi, N. {\etal} 1995;Planetary \& Space Science;43;1459
% MANDOLESI N; BERSANELLI M; CESARSKY C; DANESE L; and others.
% COBRAS/SAMBA - THE ESA MEDIUM SIZE MISSION FOR MEASUREMENTS OF CBR
% ANISOTROPY.
% PLANETARY AND SPACE SCIENCE, 1995 OCT-NOV, V43 N10-1:1459-1465.

\rf Mather, J. C. {\etal} 1994;ApJ;420;439
% MEASUREMENT OF THE COSMIC MICROWAVE BACKGROUND SPECTRUM BY
% THE COBE FIRAS INSTRUMENT.

\rf Neugebauer, G. {\etal} 1984;ApJ;278;L1

\rf O'Sullivan, C. {\etal} 1995;MNRAS;274;861
%\rn Wheelock {\etal} 1991, %{\it IRAS Sky Survey Atlas Explanatory Supplement}, IPAC.

% \rf Peebles, P. J. E. 1973;ApJ;185;413

\rf Reach, W. T. {\etal} 1995;ApJ;451;188
% preprint astro-ph/9504056.
% Title: FAR-INFRARED SPECTRAL OBSERVATIONS OF THE GALAXY BY COBE
% Author(s): W.T. Reach , E. Dwek , D.J. Fixsen , T. Hewagama , J.C. Mather , R.A. Shafer , A.J.
% Banday , C.L. Bennett , E.S. Cheng , R.E. Eplee, Jr. , D. Leisawitz , P.M. Lubin , S.M. Read , L.P.
% Rosen , F.G.D. Shuman , G.F. Smoot , T.J. Sodroski , E.L. Wright 

\rf Rybicki, G. B. \& Press, W. H. 1992;ApJ;398;169
% Rybicki, G.B.; Press, W.H.
% Interpolation, realization, and reconstruction of noisy, irregularly
% sampled data.
% Astrophysical Journal, 10 Oct. 1992, vol.398, (no.1, pt.1):169-76.
% ONE-DIMENSIONAL CASE ONLY.
% They also have a 1995 PRL where they show that you can make stuff 
% tridiagonal if the 1D power spectrum is Lorentzian.

\rf Scott D, Silk J and White M 1995;Science;268;829
% astro-ph/9505015
% SCOTT D; SILK J; WHITE M.
% FROM MICROWAVE ANISOTROPIES TO COSMOLOGY.
% SCIENCE, 1995 MAY 12, V268 N5212:829-835.

\rf Smoot, G.F. {\etal} 1992;ApJ;396;L1
% \rf Smoot, G.F.; Bennett, C.L.; Kogut, A.; Wright, E.L.; and others.
% Structure in the COBE Differential Microwave Radiometer first-year maps.

\rf Tegmark, M. \& Bunn, E. F. 1995;ApJ;455;1
% brute.tex

% \rf Smoot, G.F.; Bennett, C.L.; Kogut, A.; Wright, E.L.; and others.
% Structure in the COBE Differential Microwave Radiometer first-year maps.

\rf Tegmark, M. 1996;MNRAS;280;299
% TEGMARK M.
% A METHOD FOR EXTRACTING MAXIMUM RESOLUTION POWER SPECTRA FROM MICROWAVE SKY MAPS.
% MONTHLY NOTICES OF THE ROYAL ASTRONOMICAL SOCIETY, 1996 MAY 1, V280
% N1:299-308.
% Preprint astro-ph/9412064.
% window.tex

\rn Toffolatti, L. {\etal} 1995, in {\it 1993 Capri Workshop on the cosmic
microwave background}, 
{\it Astrophys. Lett \& Comm.}, in press. 

\rf White, M. \& Srednicki, M, 1995;ApJ;443;6

\rf White, M., Scott, D. and Silk, J. 1994;ARA\&A;32;319
% WHITE M; SCOTT D; SILK J.
% ANISOTROPIES IN THE COSMIC MICROWAVE BACKGROUND.
% ANNUAL REVIEW OF ASTRONOMY AND ASTROPHYSICS, 1994, V32:319-370.

\rf Wright, E. L. {\etal} 1991;ApJ;381;200
% WRIGHT EL; MATHER JC; BENNETT CL; CHENG ES; and others.
% PRELIMINARY SPECTRAL OBSERVATIONS OF THE GALAXY WITH A 7-DEGREES BEAM BY
% THE COSMIC BACKGROUND EXPLORER (COBE)
% ASTROPHYSICAL JOURNAL, 1991 NOV 1, V381 N1:200-209.

%\rf Wright, E. L. {\etal} 1994;ApJ;420;1
% The B_l paper.

\rf Zaroubi, S. {\etal} 1995;ApJ;449;446 
% preprint astro/ph 9410080.
% ZAROUBI S; HOFFMAN Y; FISHER KB; LAHAV O.
% WIENER RECONSTRUCTION OF THE LARGE-SCALE STRUCTURE.
% ASTROPHYSICAL JOURNAL, 1995 AUG 20, V449 N2:446-459.

%%%%%%%%%%%%%%%%%%%%%% FIGURES: %%%%%%%%%%%%%%%%%%%%%%%%%

\clearpage

\newpage

\newpage

\newpage

\newpage
\begin{figure}[phbt]
\centerline{\epsfxsize=17cm\epsfbox{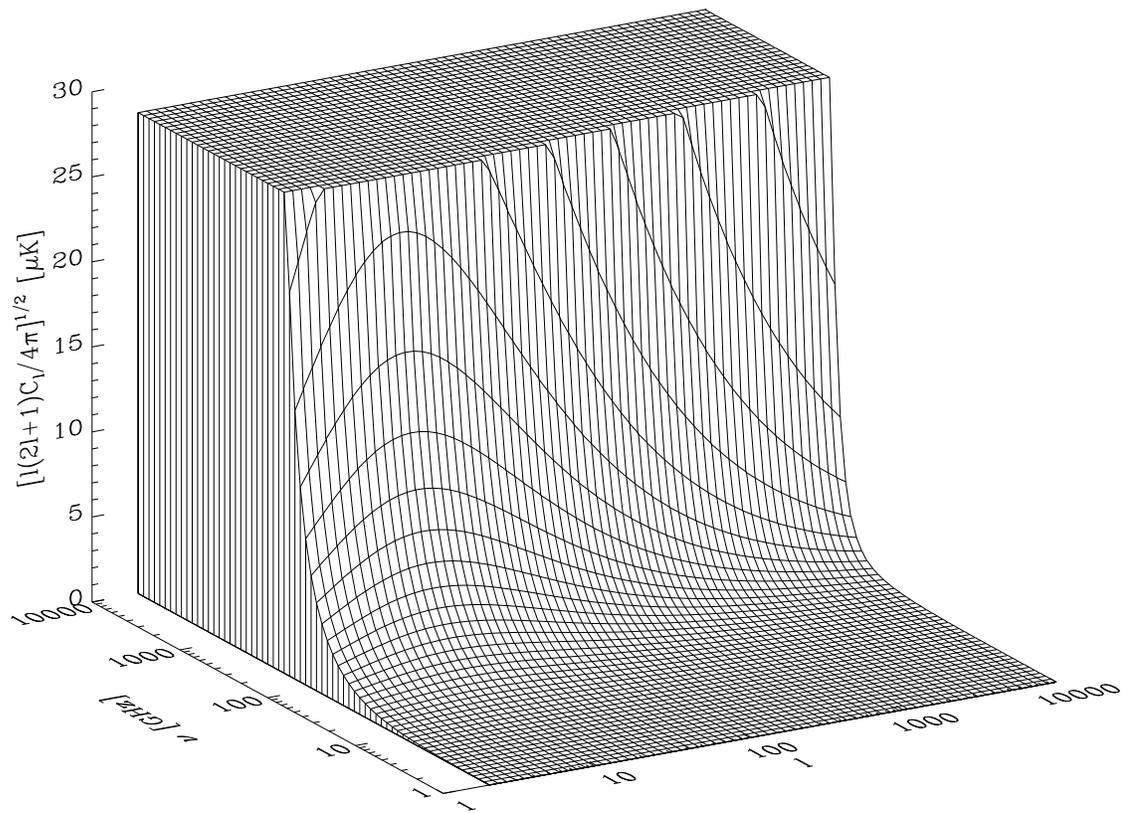}}
\caption{Model for how the dust power spectrum
depends on multipole and frequency.}
\label{DustFig}
\end{figure}

\newpage
\begin{figure}[phbt]
\centerline{\epsfxsize=17cm\epsfbox{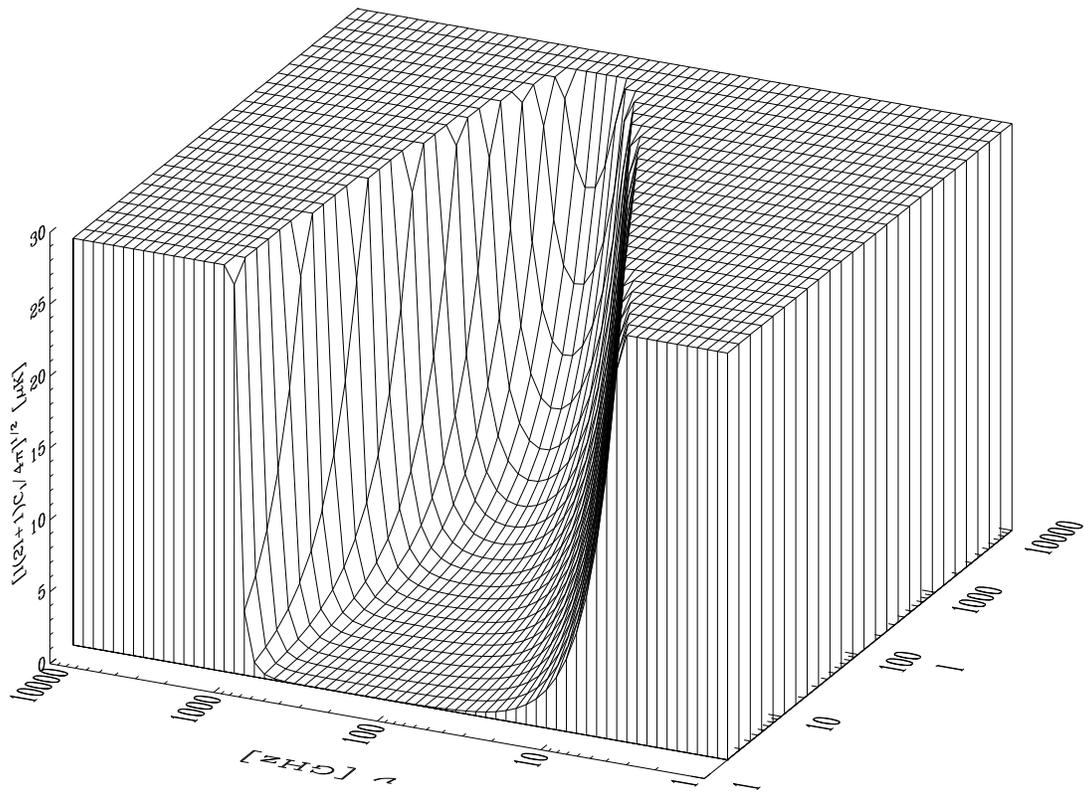}}
\caption{Model for how the power spectrum of 
point sources depends on multipole and frequency.}
\label{pointsourcesFig}
\end{figure}

\newpage
\begin{figure}[phbt]
\centerline{\epsfxsize=17cm\epsfbox{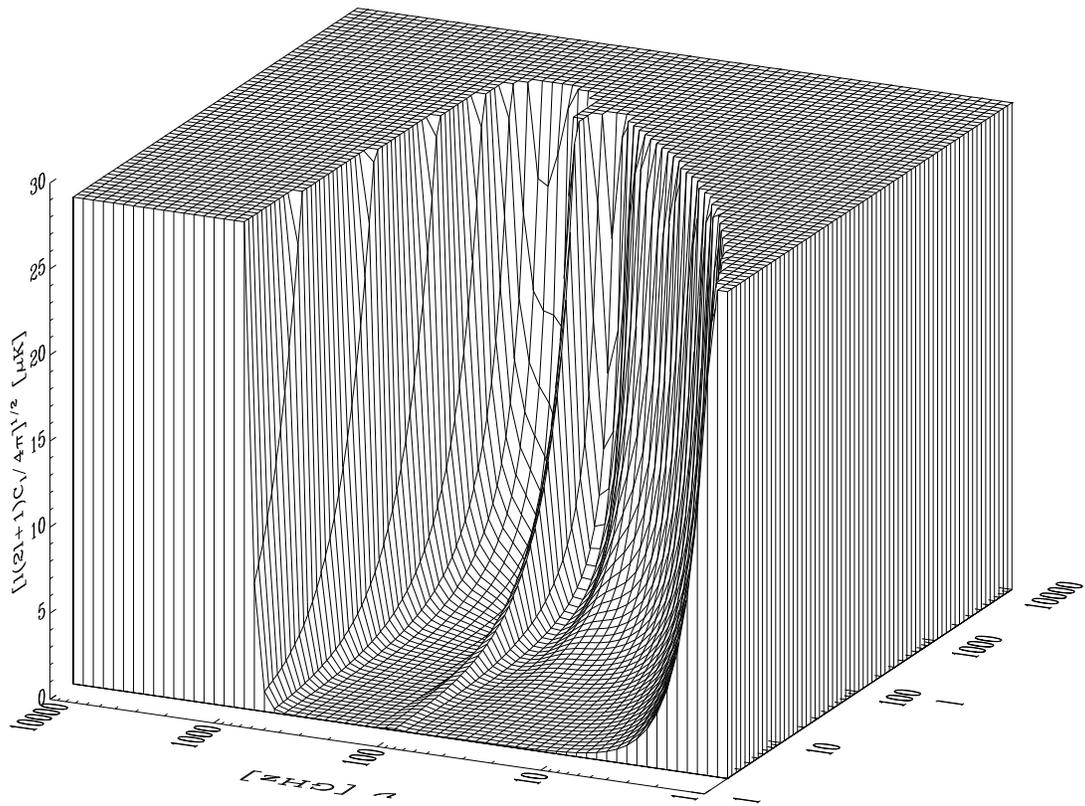}}
\caption{How the COBRAS/SAMBA pixel noise 
power depends on multipole and frequency.}
\label{NoiseFig}
\end{figure}

\newpage
\begin{figure}[phbt]
\centerline{\epsfxsize=17cm\epsfbox{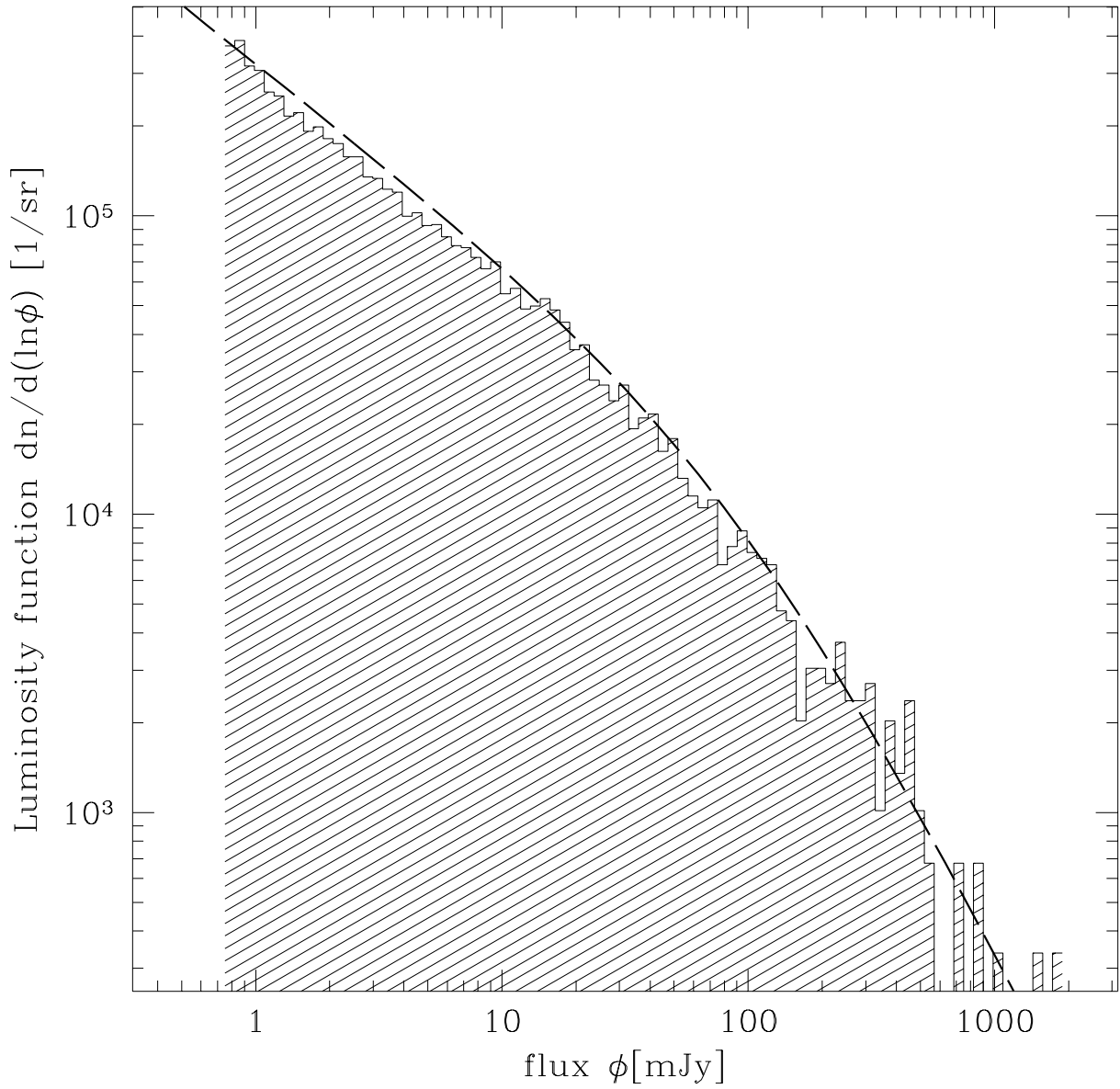}}
\caption{The VLA FIRST 
differential source count
for radio sources at 1.5 GHz.}
\label{FluxFig}
\end{figure}
 
\newpage
\begin{figure}[phbt]
\centerline{\epsfxsize=17cm\epsfbox{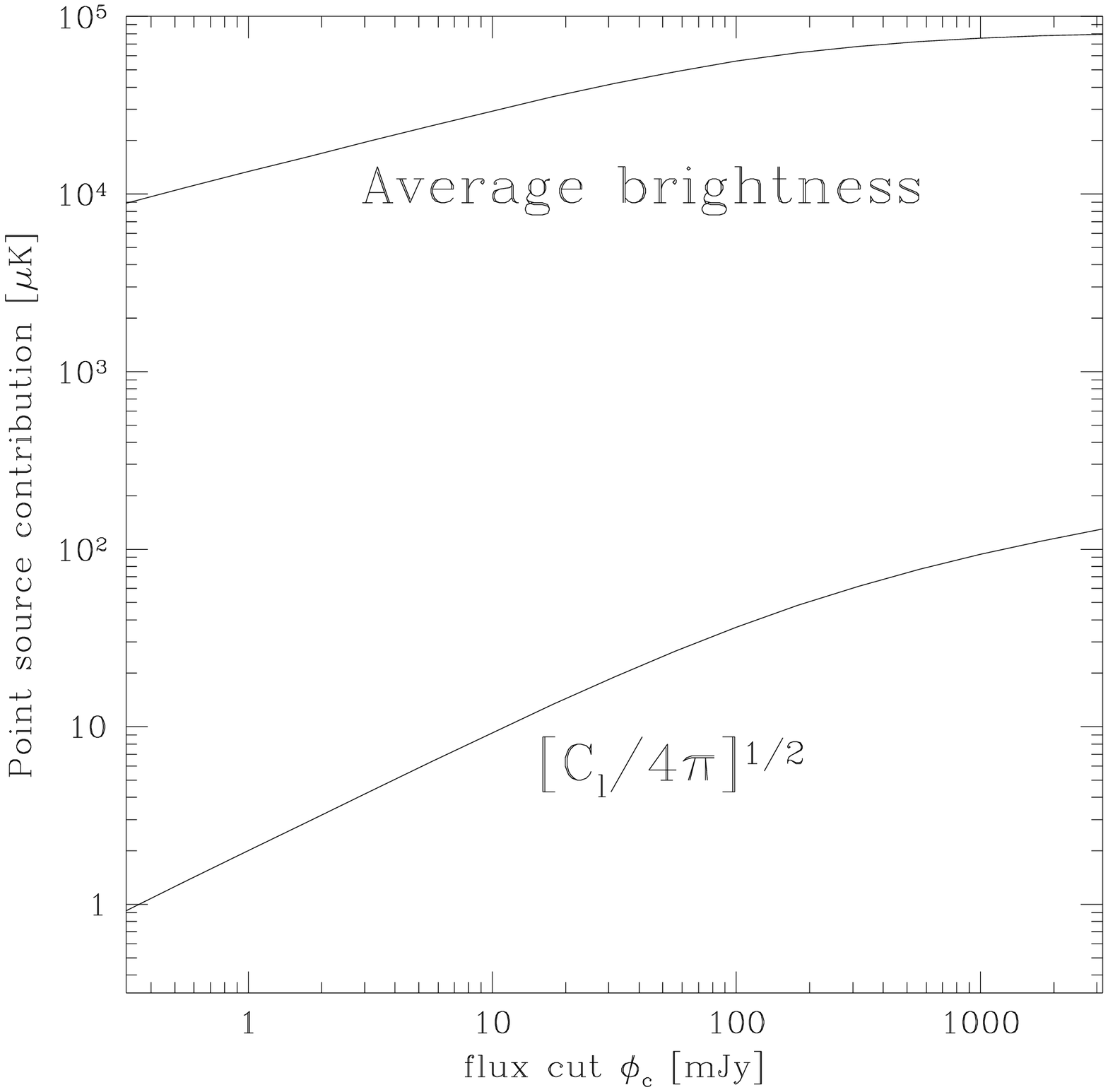}}
\caption{Dependence of radio source fluctuations on flux cut.}
\label{RadioClFig}
The average brightness (monopole, upper curve) and the power spectrum
normalization (lower curve) at 1.5 GHz 
from the VLA FIRST survey is plotted 
as a function of flux cut.
\end{figure}

\newpage
\begin{figure}[phbt]
\centerline{\epsfxsize=17cm\epsfbox{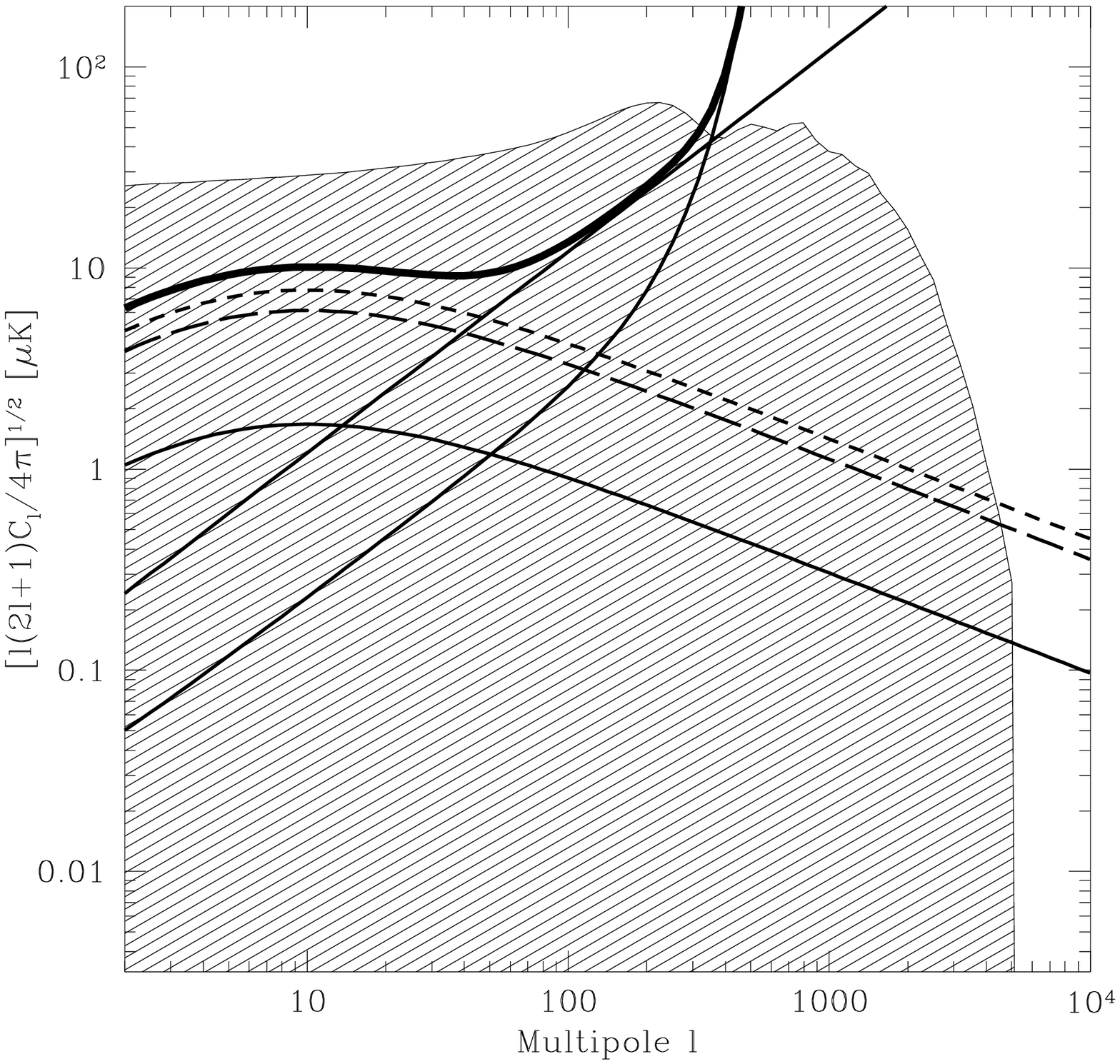}}
\caption{Model power spectra of various components at 31 GHz.} From 
the bottom up, on the left hand side, the curves correspond to
pixel noise, radio point sources, dust, synchrotron radiation, 
free-free emission, all sources combined (heavy), and CMB (shaded)
according to the CDM model plotted in Figures 2 and 3.
\label{Channel31Fig}
\end{figure}

\newpage
\begin{figure}[phbt]
\centerline{\epsfxsize=17cm\epsfbox{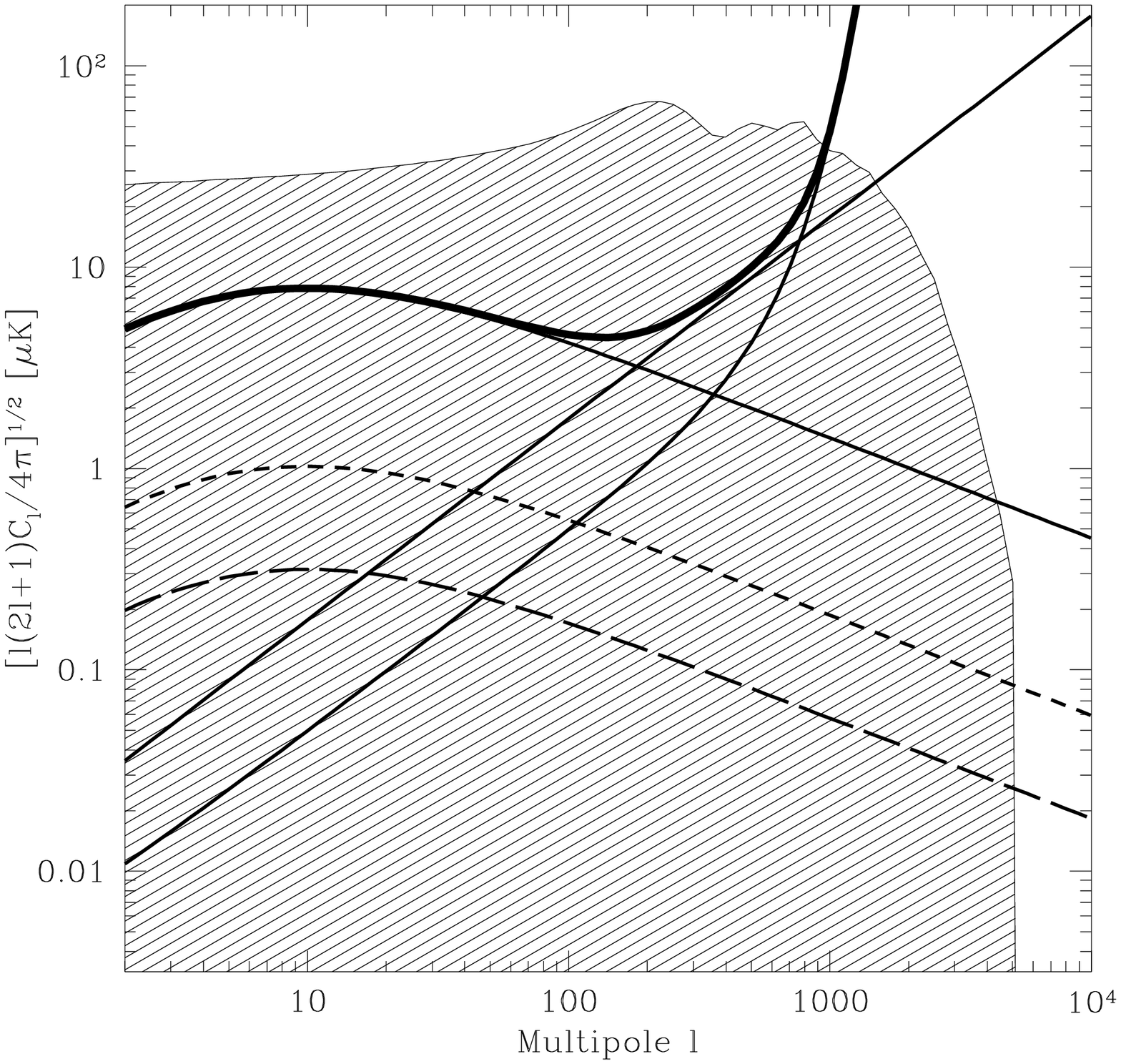}}
\caption{Model power spectra of various components at 90 GHz.} From 
the bottom up, on the left hand side, the curves correspond to
pixel noise, radio point sources, synchrotron radiation, 
free-free emission, dust, all sources combined (heavy), 
and CMB (shaded).
\label{Channel90Fig}
\end{figure}

\newpage
\begin{figure}[phbt]
\centerline{\epsfxsize=17cm\epsfbox{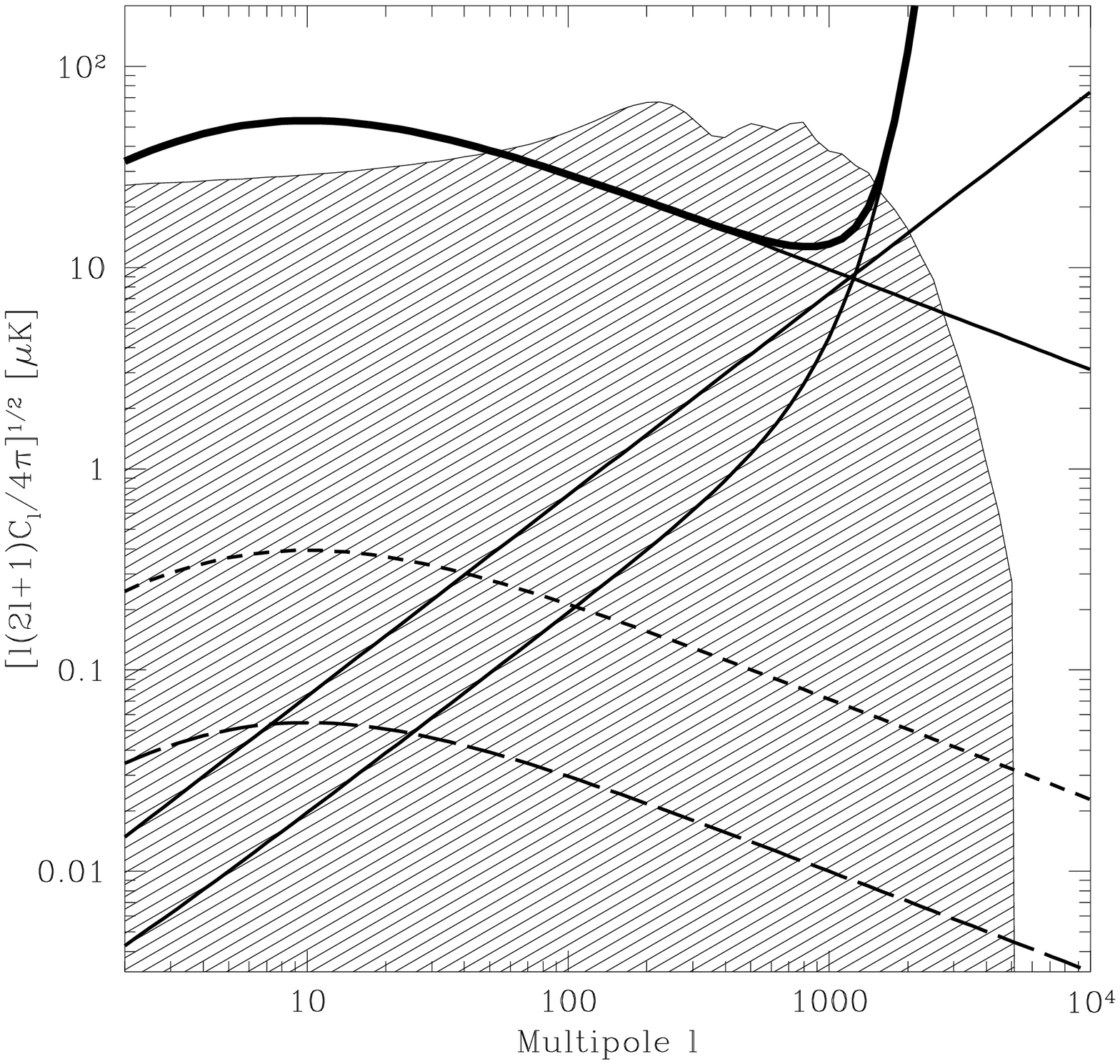}}
\caption{Model power spectra of various components at 217 GHz.} From 
the bottom up, on the left hand side, the curves correspond to
pixel noise,
radio point sources, synchrotron radiation, 
free-free emission, CMB (shaded), dust, 
and all sources combined (heavy).
\label{Channel217Fig}
\end{figure}

\newpage
\begin{figure}[phbt]
\centerline{\epsfxsize=17cm\epsfbox{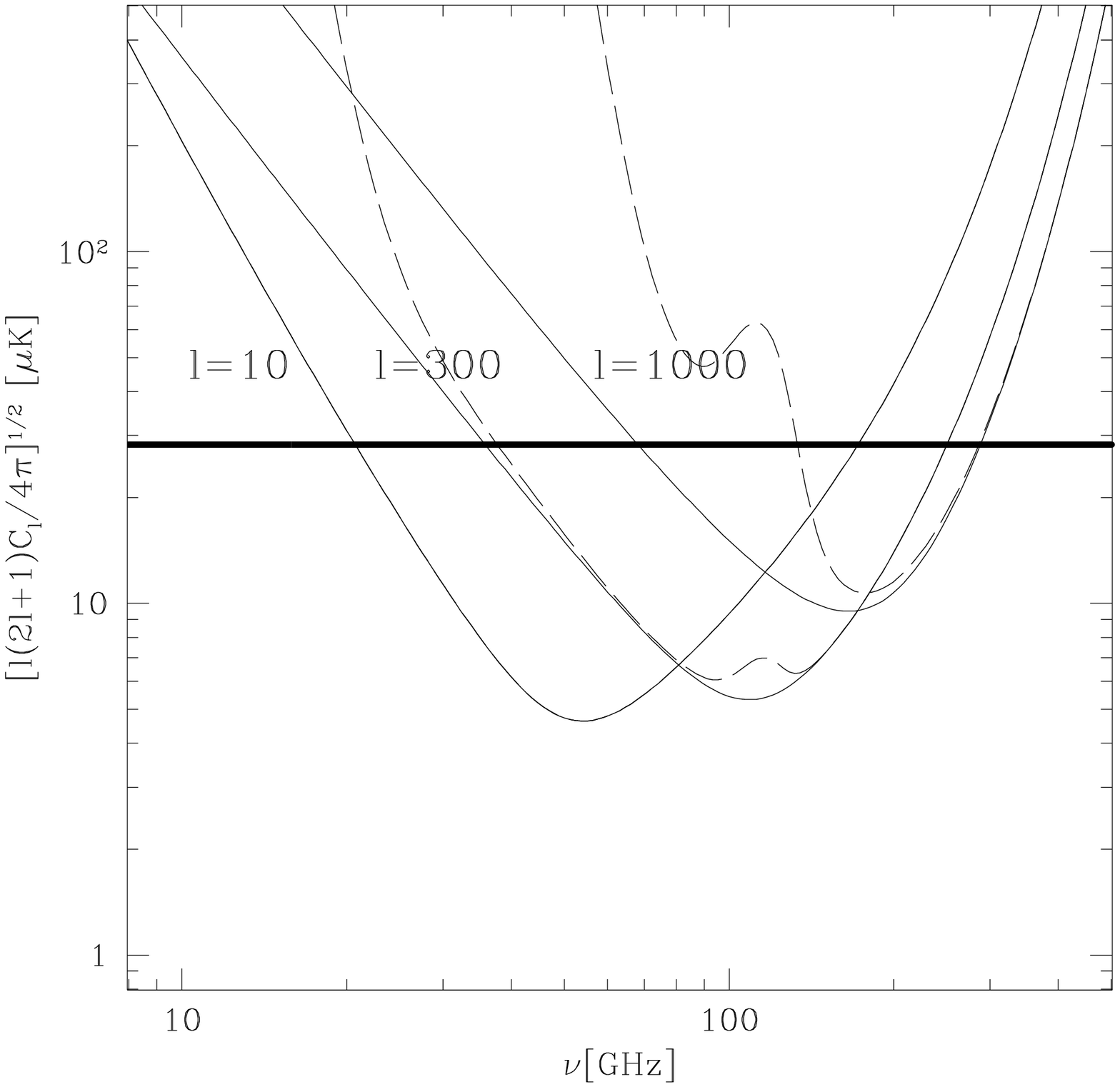}}
\caption{Total contamination at selected
multipoles.}
The total fluctuations of all foregrounds excluding (solid lines) and including
(dashed lines) COBRAS/SAMBA pixel noise are plotted for 
the multipoles 10, 300 and 1000. The heavy horizontal line corresponds to
COBE-normalized Sachs-Wolfe fluctuations in the CMB.
\label{MultipolesFig}
\end{figure}

\newpage
\begin{figure}[phbt]
\centerline{\epsfxsize=17cm\epsfbox{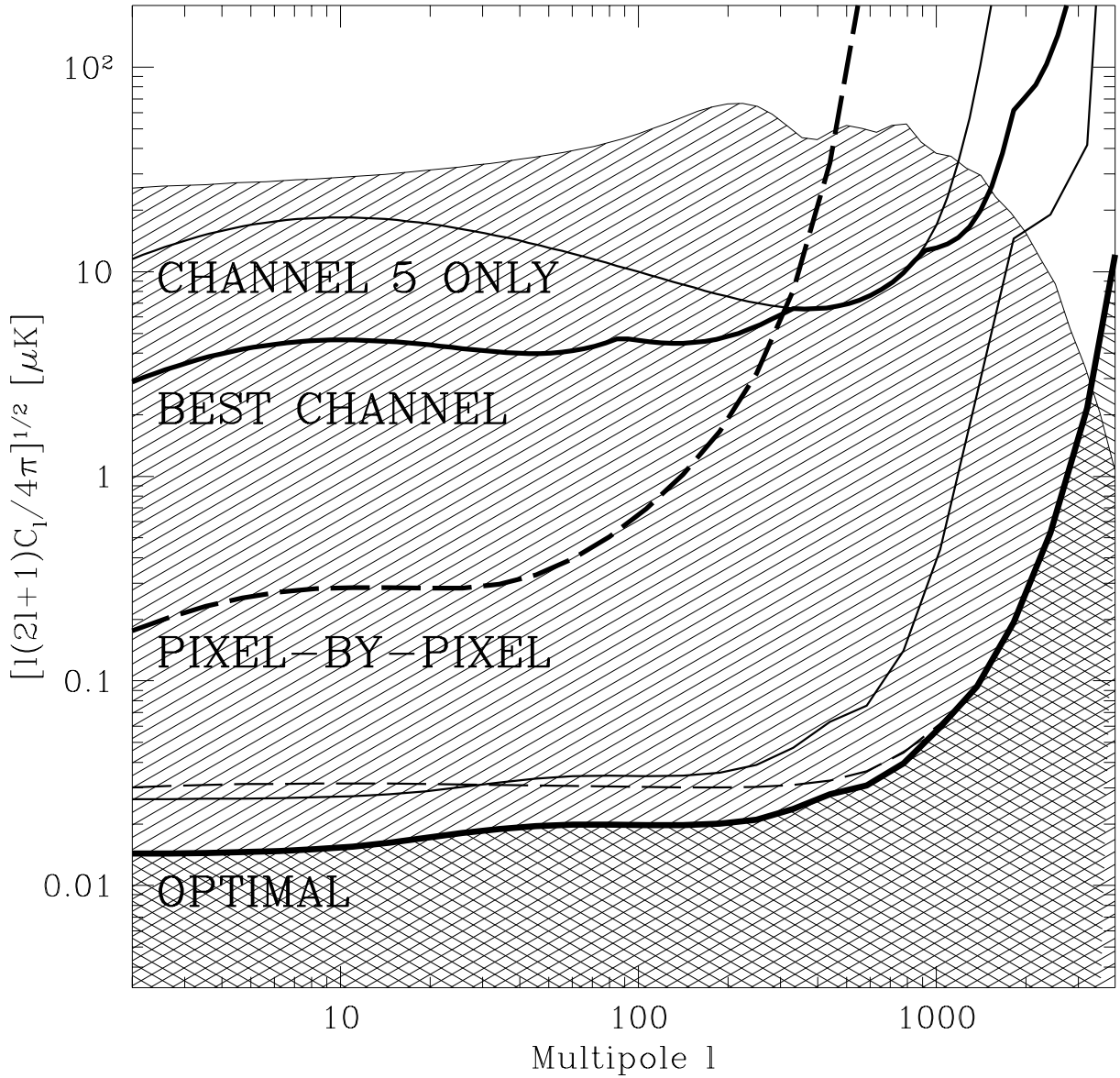}}
\caption{Comparison of methods}
The combined residual contribution of all foregrounds and noise 
is plotted
for different approaches to foreground subtraction. From 
top to bottom, the four labeled curves correspond to
(1) use of the 143 GHz channel with no subtraction, (2) use of the best
COBRAS/SAMBA channel at each multipole with no subtraction, 
(3) optimized subtraction on a pixel-by-pixel basis, and
(4) the optimal subtraction technique.
The uppermost curve (shaded) is a standard CDM power spectrum
as plotted in earlier figures.
The two thin curves at the bottom correspond to reducing 
the number of
channels as described in the text.
\label{MethodsFig}
\end{figure}

\newpage
\begin{figure}[phbt]
\centerline{\epsfxsize=17cm\epsfbox{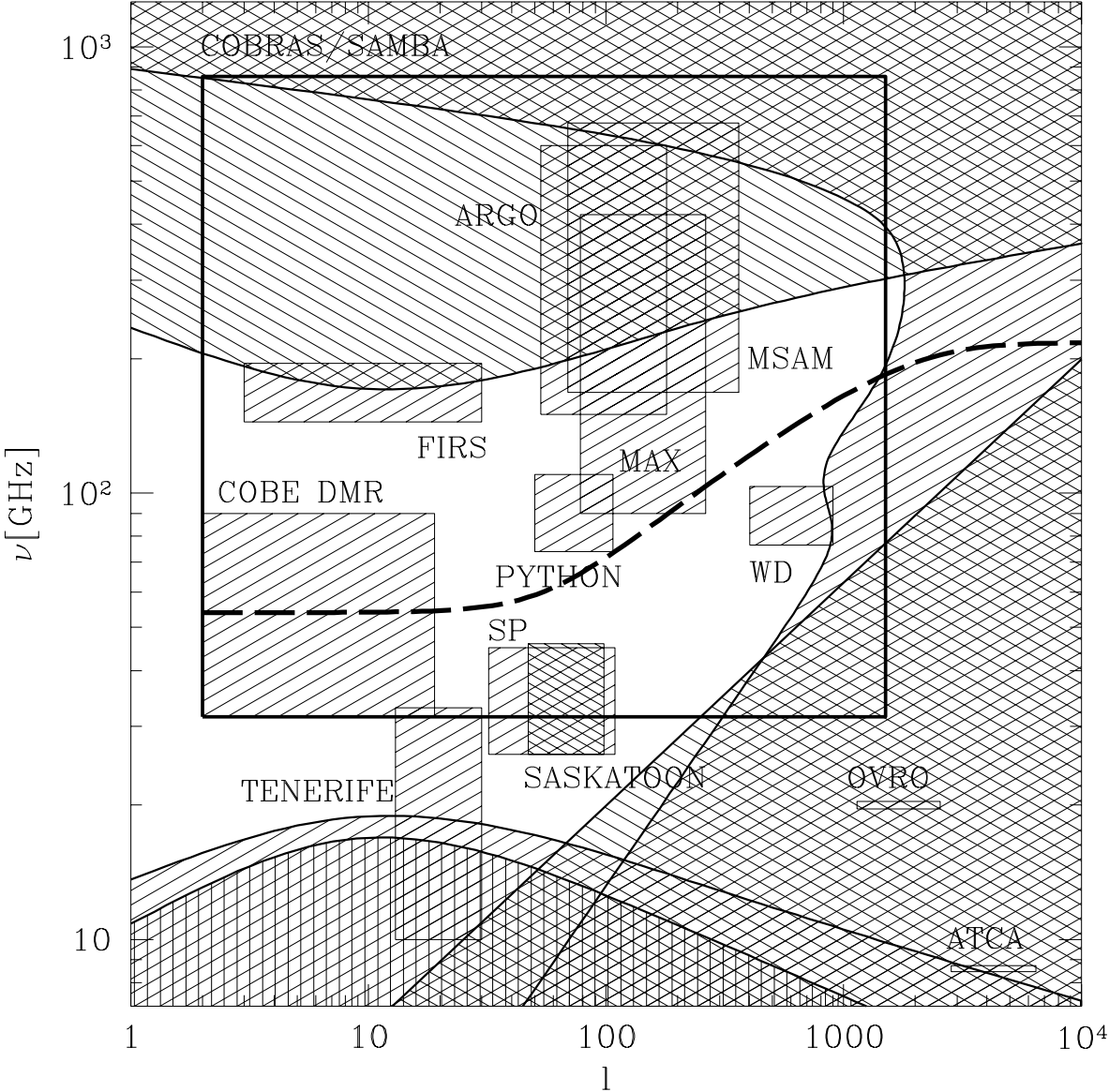}}
\caption{Where various foregrounds dominate}
The shaded regions indicate where the various foregrounds cause 
fluctuations exceeding those of COBE-normalized scale-invariant
fluctuations ($\approx\sqrt{2}\times 20\muK$), 
thus posing a substantial challenge
to estimation of genuine CMB fluctuations. 
They correspond to dust (top), free-free emission (lower left), 
synchrotron radiation (lower left, vertically shaded),
radio point sources (lower right) and COBRAS/SAMBA 
instrumental noise and beam dilution (right).
The heavy dashed line shows the frequency where the total foreground
contribution to each multipole is minimal.
The boxes indicate roughly the range of multipoles $\l$ and frequencies
$\nu$ probed by various CMB experiments, as in \fig{ExperimentsFig}.
\label{EverythingFig}
\end{figure}

\end{document}